\documentclass{IEEEtran}
\usepackage{cite}
\usepackage{amsmath,amssymb,amsfonts}
\newtheorem{rem}{Remark}
\usepackage{graphicx}
\usepackage{textcomp}
\usepackage{xcolor,soul,framed} 
\usepackage{multirow}    
\usepackage{multicol}         
\usepackage{bbding}
\usepackage{subfigure}
\usepackage{tikz}
\usetikzlibrary{spy}
\usepackage{bm}
\usepackage{pgfplots}

\def\BibTeX{{\rm B\kern-.05em{\sc i\kern-.025em b}\kern-.08em
    T\kern-.1667em\lower.7ex\hbox{E}\kern-.125emX}}

\usepackage{algorithm}
\usepackage{algpseudocode}
\usepackage{booktabs}
\usepackage{tabularx}
\usepackage{makecell}
\definecolor{mycolor5}{rgb}{0.635, 0.078, 0.184}  

\ifCLASSINFOpdf
 
\else

\fi

\begin{document}

\title{Propagation Mechanism-Aware Near-Field Spatially Non-Stationary Channel Estimation and Environment Mapping}
\author{
Yuan~Liu,~\IEEEmembership{Member,~IEEE,}
Xuesong~Cai,~\IEEEmembership{Senior~Member,~IEEE,}
Dipankar~Saha,~\IEEEmembership{Student~Member,~IEEE,}
M. R. Bhavani Shankar,~\IEEEmembership{Senior~Member,~IEEE,}
Björn Ottersten~\IEEEmembership{Fellow,~IEEE}
\thanks{Y. Liu, D. Saha, B. Shankar, and B. Ottersten are with the Interdisciplinary
Centre for Security, Reliability and Trust (SnT), University of Luxembourg, L-1855, Luxembourg (e-mail: \{yuan.liu, dipankar.saha, bhavani.shankar, bjorn.ottersten\}@uni.lu). 
}
\thanks{X. Cai is with the School of Electronics, Peking University,
Beijing, 100871, China (email: xuesong.cai@pku.edu.cn).}
}

\maketitle
\begin{abstract}
Extremely large aperture arrays (ELAAs) benefit the dual functions of integrated sensing and communication (ISAC) systems by enabling high-throughput data streams and high angular resolution with near-field spatial diversity. 
However, near-field spherical wavefront effects and spatial non-stationarity (SNS) bring challenges to both communication and sensing. 
This paper studies near-field spatially non-stationary channel estimation and environment mapping by jointly accounting for multi-bounce, blockage-induced partial visibility, and hybrid reflection-scattering propagation.
We propose a unified parametric sensing channel model that represents the SNS phenomenon (due to partial array blockage, diffraction, and specular reflection) through spatially varying visibility and amplitude of each multipath across the array. 
To regularize the spatially varying delays caused by propagation mechanisms, we incorporate geometric constraints (GCs) based on environmental interaction points, embedding them into the model as absolute propagation delays.
We then develop a GC-space-alternating generalized expectation-maximization (GC-SAGE) algorithm to estimate near-field channel parameters and locate environment scatterers/reflectors.
Moreover, the GC-SAGE calculates per-antenna path amplitudes based on the delays determined by the coordinates of scatterers/reflectors and transceivers, thereby effectively detecting channel SNS.
Both ray-based simulation and field measurement are used to validate the proposed approach.
\end{abstract}

\begin{IEEEkeywords}
Environment mapping, extremely large antenna aperture array (ELAA), GC-SAGE, near-field, scatterer and reflector localization, spatial non-stationarity (SNS).
\end{IEEEkeywords}
\section{Introduction}
Integrated sensing and communications (ISAC) has emerged as a key enabling technology for the sixth-generation (6G) wireless networks, which is expected to support a wide range of Internet-of-Things (IoT) applications \cite{mishra2024signal,9755276,11126933,witrisal2016high,liu2025doppler,11113304}.
In short-range IoT scenarios, the propagation channel is strongly shaped by the surrounding environment. 
On the one hand, such complex propagation is challenging to reliable channel estimation and thereby degrades the robustness of both communication and sensing.
On the other hand, environment-induced propagation mechanisms inherently encode signatures of the surrounding scene, since each multipath component is geometry-dependent and can be attributed to one or multiple scatterers/reflectors in the environment \cite{9215972,9938129}. 
Therefore, it is crucial to account for realistic propagation mechanisms in model-based parameter estimation and environment-aware applications.

\subsection{Near-Field and Spatially Non-Stationary Channel Model}
To enable highly directive beamforming and spatial multiplexing while improving range and angular resolutions, extremely large aperture arrays (ELAAs) have been widely investigated in millimeter-wave (mmWave) and beyond ISAC systems \cite{10445208,8828030,11183598}.
As the array aperture grows, the near-field effect becomes non-negligible in ELAA systems\footnote{In wireless channels, we are not referring to the reactive near-field of antennas. Instead, the term near-field typically denotes the radiating near-field (Fresnel region), i.e., the distance range $[0.62\sqrt{{D^3}/{\lambda}}, {2D^2}/{\lambda}]$, where $D$ and $\lambda$ denote the aperture size and the wavelength, respectively.}.
Near-field channel modeling was already explored as early as 2005 in \cite{1510955}, where a spherical wavefront formulation was adopted. 
%
Subsequently, the spherical wavefront-based model has been extensively investigated across wireless communications, estimation, and localizations \cite{lu2024tutorial,channelestimationxlmimo,7501567}. 

Besides near-field effects, another conventional channel phenomenon, namely spatial non-stationarity (SNS), has also attracted renewed attention in ELAA systems. 
In early works, SNS refers to the spatial variation of multipath composition and channel statistics observed by spatially separated antenna elements, and it is commonly characterized by the visibility region of each cluster \cite{Liu2012COST2100}.
Later, for large-scale distributed antenna deployments, SNS is characterized by modeling cluster evolution through \textit{birth to death} processes \cite{Wu2015NonStationaryMassiveMIMO}.
In near-field measurements with ELAAs, element-dependent per-path variations become more pronounced, since different parts of the array may experience unresolved multipath components, partial blockage, and imperfect hardware coupling \cite{8713575}. 
As a result, recent SNS channel modeling studies have associated geometry-dependent variations across ELAAs with environment-induced propagation mechanisms, which indicates that SNS is mainly caused by inhomogeneous scattering, including specular reflection and diffraction in blockage scenarios \cite{10179246,9940939,JHZhang2026jsac_fr3_xlmimo}.

%

\subsection{Channel Parameter Estimation and Mapping}
Environment sensing-oriented tasks often rely on high-resolution channel parameter estimation to extract per-path parameters such as delay, angle of arrival (AoA), and angle of departure (AoD), which can then be associated with physical objects and structures in the scene. 
Representative approaches include the space-alternating generalized expectation-maximization (SAGE) \cite{fleury1999channel}, the RiMAX framework for multidimensional channel sounding \cite{1351040}, and orthogonal matching pursuit (OMP)-based sparse reconstruction \cite{Tropp2007OMP}.
Due to the multi-bounce nature of radio propagation, a multipath component may traverse multiple scatterers and reflectors before reaching the receiver (Rx) \cite{ling2017experimental}. Accordingly, several environment mapping works have investigated multi-bounce propagation and incorporated it into their sensing and mapping models \cite{leitinger2023datafusion,10694028,feng2024multipathghost_trs}. 
In near-field, the spherical wavefront is essentially utilized as a range-angle dependent array response\footnote{Since Doppler stems from the time derivative of the range, this coupling is also extended to range-velocity-angle dependent array response \cite{Wei2025FundamentalLimitsNFPartII}.} \cite{10964143,9598863}.
In environment mappings, the spherical model provides additional geometric constraints (GCs) for multi-bounce propagation, enabling the localization of the first-hop and last-hop interaction points associated with the transmitter (Tx) and Rx arrays \cite{7981398}. 
Leveraging this feature, several SAGE-family algorithms have been developed for multi-bounce multipath parameter estimation, scatterer localization, and environment mapping \cite{Yuan_TWC25,9938129,10038714}.
It is worth mentioning that locating high-bounce (more than two-bounce) paths remains ambiguous in bistatic settings, as discussed in \cite{Yuan_TWC25}. 

As SNS becomes increasingly pronounced in ELAA systems, the combined effects of near-field propagation and SNS have already been explored in measurement-driven channel modeling studies~\cite{10179246,9940939,JHZhang2026jsac_fr3_xlmimo}. 
However, comprehensive channel estimation studies remain limited. 
Recent works on localization and environment mapping have begun to investigate certain individual SNS-related effects, such as partial blockage \cite{10447918,liu2025environment} and diffraction-aided sensing \cite{duggal2025diffraction,10942642}.
In comparison, sensing-oriented formulations for reflection-induced SNS channels remain underexplored.
This issue is particularly important because, as noted in \cite{7981398}, adopting scattering-only models to hybrid reflection-scattering channels may produce mirror-type ghost estimates in the reconstructed geometry. Nevertheless, scattering-only assumptions are still widely used in radio-based sensing and mapping studies \cite{7501567,leitinger2023datafusion,10694028,9827865,Yuan_TWC25,10038714,7981398}. 
%
Therefore, developing a unified mechanism-aware channel estimation framework for sensing and environment mapping that accounts for SNS effects due to partial blockage, hybrid reflection-scattering propagation, and diffraction remains an open problem.

\subsection{Contribution}
%
This paper develops a propagation-induced GC-SAGE algorithm that derives GCs from physical propagation mechanisms and incorporates them into joint near-field SNS channel estimation and environment mapping.
The main contributions are summarized as follows:
\begin{itemize}
  \item \textbf{Unified parametric near-field SNS channel model:}
  We propose a unified parametric near-field channel model, which jointly accounts for multi-bounce propagation and SNS caused by partial array illumination under different propagation mechanisms, including blockage, reflection, and diffraction.
    \item \textbf{Geometry constraints for interaction points:} 
    Each propagation path is represented as a sequence of interaction points, which link per-path channel parameters to geometry-dependent constraints between ELAAs and coordinates of scatterers/reflectors. 
    Scattering is coherent to the whole array, whereas specular reflection yields element-dependent points tied by a common surface normal and partial illumination. Diffraction at blockage edges is modeled as a coherent interaction with non-uniform illumination.
  
    \item \textbf{GC-SAGE algorithm:}
    Building upon the unified model and the modeled GCs, we develop a GC-SAGE algorithm for joint near-field SNS channel parameter estimation and environment mapping. Specifically, it localizes reflecting surfaces and scatterers, accounts for mechanism-induced element-dependent partial array illumination.

  \item \textbf{Measurement-based validations:}
  The unified sensing channel model and GC-SAGE are validated 
  via ray-tracing (RT) simulations and indoor measurements under both monostatic and bistatic settings, covering partial blockage, multi-bounce, hybrid reflection-scattering, LoS, and OLoS scenarios.
  Results confirm the accuracy of the near-field SNS model and the effectiveness of GC-SAGE in channel estimation and environment mapping. 
\end{itemize}

\textbf{Notations:}
The following notations are used throughout this paper.
Bold lower-case and upper-case letters denote vectors and matrices, respectively.
In particular, $\mathbf{A} \in \mathbb{R}^{N_1 \times N_2}$ and $\mathbf{A} \in \mathbb{C}^{N_1 \times N_2}$ denote real-valued and complex-valued matrices of size $N_1 \times N_2$, respectively, while $\mathbf{a} \in \mathbb{R}^{N_1}$ and $\mathbf{a} \in \mathbb{C}^{N_1}$ denote real-valued and complex-valued column vectors of length $N_1$, respectively.
$\mathbf{I}$ denotes the identity matrix.
$(\cdot)^T$ and $(\cdot)^H$ denote the transpose and conjugate transpose, respectively.
$\|\cdot\|_0$ and $\|\cdot\|_p$ denote the $\ell_0$ and $\ell_p$ norms, respectively.
In particular, $|\mathbf{a}|$ denotes its Euclidean norm, i.e., $|\mathbf{a}|=\|\mathbf{a}\|_2$, whereas for a scalar $a$, $|a|$ denotes its absolute value.
$\otimes$ denotes the Kronecker product, and $\odot$ denotes the element-wise (Hadamard) product.

The rest of the paper is organized as follows. 
Section~\ref{Sec:signal_model} introduces the unified parametric model, GCs of different propagation mechanisms, and the problem formulation. 
Section~\ref{Sec:Algo} presents the proposed GC-SAGE algorithm. 
Section~\ref{sec:simu} provides details on both simulation and measurement-based validations. 
Section~\ref{sec:conclude} concludes the paper.

\begin{figure*}[ht]
    \centering
    \includegraphics[width=0.95\linewidth, height=0.48\linewidth]{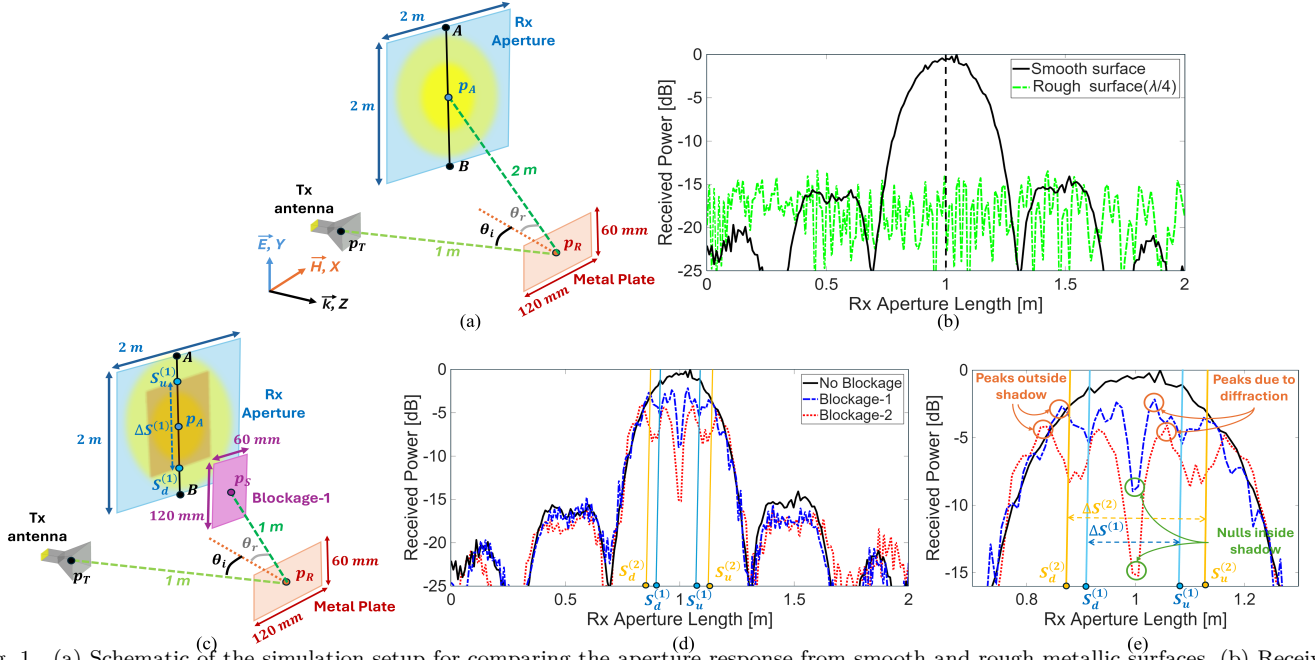}
    \vspace{-0.5cm}
    \caption{
    (a) Schematic of the simulation setup for comparing the aperture response from smooth and rough metallic surfaces. 
    (b) Received power along the measurement line $AB$.
    (c) Schematic of the simulation setup used to evaluate diffraction effects due to different sizes of blockages. 
    (d) Received power along $AB$ under no blockage, Blockage-1, and Blockage-2, respectively.
    (e) A zoomed-in view highlighting the peak responses. 
    }
    \label{fig_Diffrection_Sim}
    \vspace{-0.5cm}
\end{figure*}
\begin{figure}[ht]
\centering
\subfigure[Reflection and scattering propagation \label{fig:paralell_normals}]
    {\includegraphics[width=0.42\textwidth]{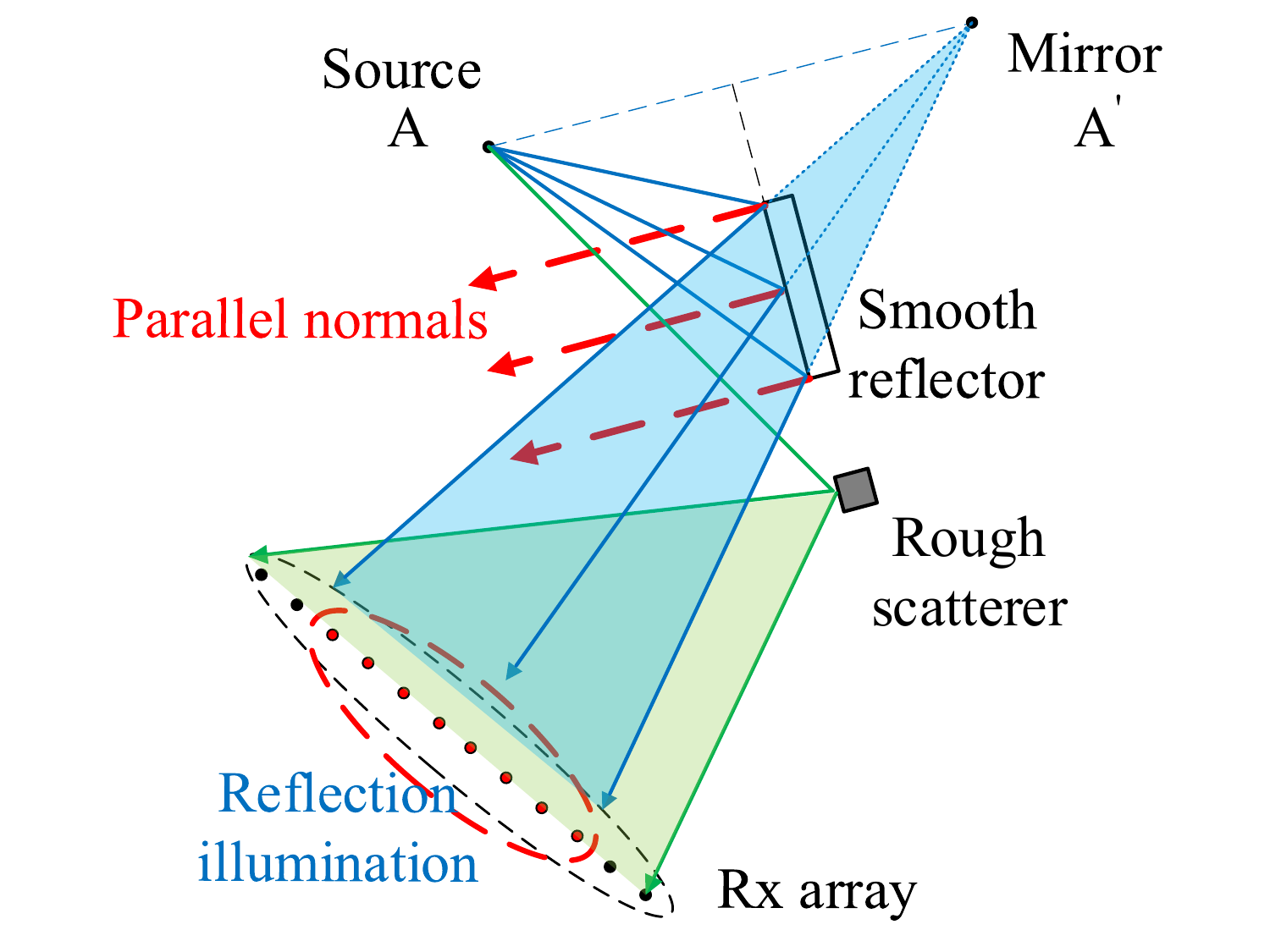}}
    \hfill
\subfigure[Partial blockage and diffraction one the edges of obstacle  \label{fig:difrraction_1}]{\includegraphics[width=0.42\textwidth]{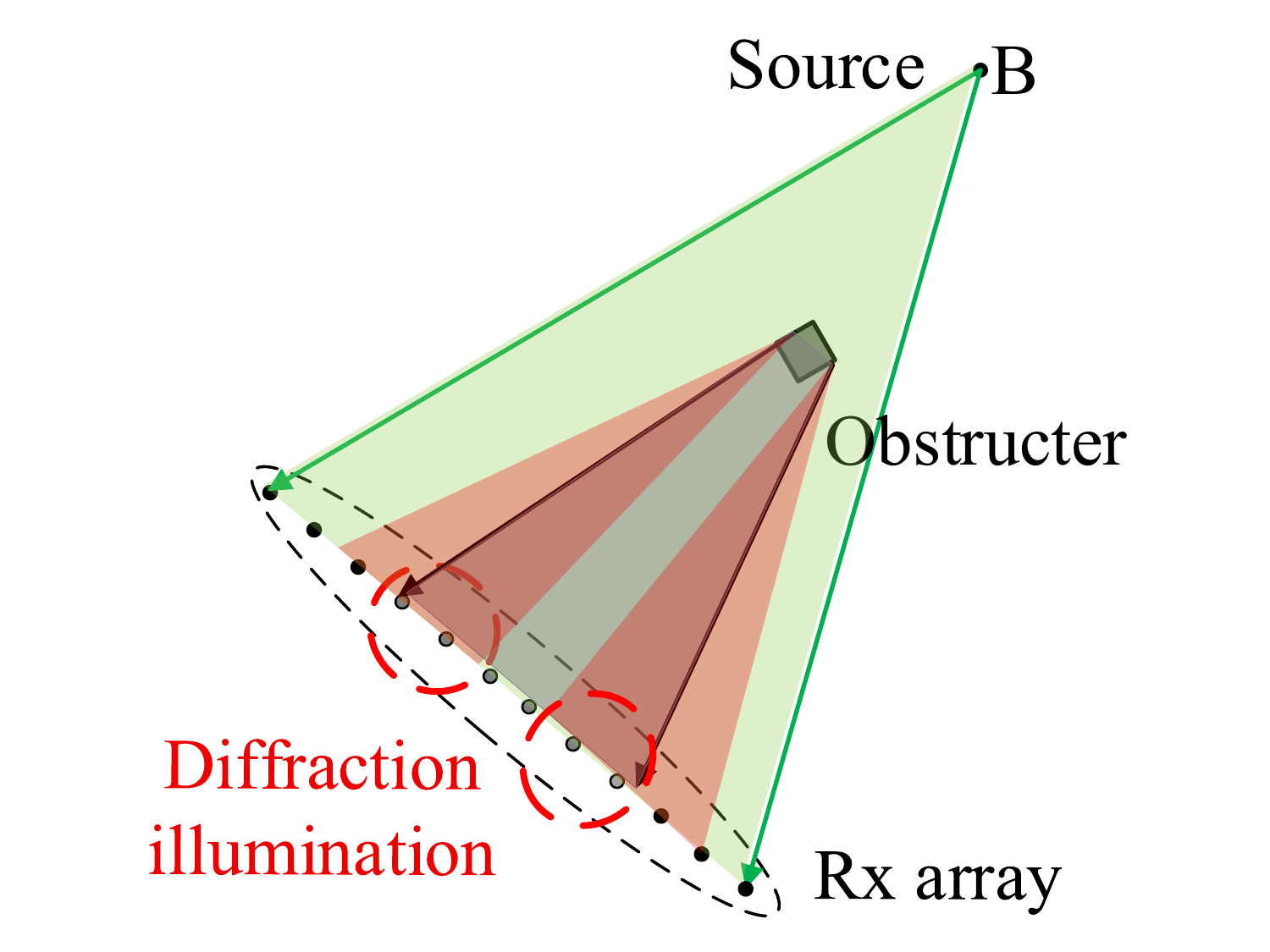}}
    \hfill
     \vspace{-0.2cm}
\caption{Geometric models of scattering, reflection, blockage, and diffraction.
}
\label{fig:propagation_mechanism}
\vspace{-0.5cm}
\end{figure}
\section{Signal model and problem formulation}\label{Sec:signal_model} 
%
In the near-field ELAA scenario of interest, the received signal may contain LoS and multi-bounce components, where multi-bounce propagation can be caused by one/more successive bounces, including reflection, scattering, blockage, and diffraction, determined by the environment geometry and wave-environment interactions.
As a result, some propagation mechanisms can give rise to SNS across the array.

To have a physically grounded parametric signal model for channel estimation and environment mapping, we provide a review of radio propagation mechanisms.
\subsection{Characterization of Representative Radio Wave Propagation Mechanisms via Full-Wave Simulations}
In this subsection, we illustrate several SNS-related representative propagation scenarios in the considered sensing setups and their induced aperture-domain responses using the full-wave electromagnetic solver (Ansys HFSS) as shown in Fig.~\ref{fig_Diffrection_Sim}. 
Our purpose is not to provide a detailed electromagnetic computation, but to identify mechanism-dependent spatial signatures that motivate the signal model developed later.

\subsubsection{Smooth and Rough Metal-Plate Reflector}
Fig.~\ref{fig_Diffrection_Sim}(a) shows the initial simulation setup which comprises: a Tx horn antenna extended by a $WR-34$ rectangular waveguide, a $60\,\text{mm}\times120\,\text{mm}$ metal-plate reflector and an Rx aperture of $2\,\text{m}\times2\,\text{m}$. The Tx horn antenna illuminates the metal plate at an incidence angle of $\theta_i=30^\circ$. The line connecting the centers of the metal plate and the Rx aperture is set to the expected outgoing angle of the reflected wave $\theta_r=30^\circ$ with respected to the normal of the metal plate based on Snell's law. To ensure proper alignment with the measurement plane, the phase-center of the horn antenna is positioned at the same height as the centers of both the metal plate and the Rx aperture. The electric field is polarized along the $Y$ axis and the measurement plane is the plane of incidence and the reflected field is observed over the Rx aperture along line $AB$. The received electric-field powers via smooth and rough metal surfaces are shown in Fig.~\ref{fig_Diffrection_Sim}(b). 

Reflection from a smooth surface results in a concentrated power distribution (solid black line) over the Rx aperture around the expected specular direction $\theta_r=30^\circ$. Due to this energy concentration, only parts of the Rx aperture receive significant power, resulting in pronounced SNS. In contrast, scattering leads to a nearly uniform power distribution across the Rx aperture, producing a less directive pattern (dotted green line) and negligible SNS. These observations indicate that \textbf{reflection behaves as a mirror-like mechanism based on geometrical optics principle, whereas scattering acts as a diffuse mechanism with spatially spread energy}. 
This distinction motivates the modeling of reflection and scattering GCs, which are illustrated in Fig.~\ref{fig:paralell_normals}.
%
\subsubsection{Diffraction and Blockage}
To explore blockage and diffraction effects, we put obstacles into the smooth-reflection scenario as shown in Fig.~\ref{fig_Diffrection_Sim}(c), where two obstacle sizes are compared, with heights of $120$~mm and $240$~mm along the $Y$-axis. 
The corresponding received electric-field powers and a zoomed view are shown in Fig.~\ref{fig_Diffrection_Sim}(d) and (e), respectively. 
 
As depicted in the schematic Fig.~\ref{fig_Diffrection_Sim}(c), a shadowed region over the Rx aperture is expected due to the presence of the obstacle, leading to partial visibility of the reflected power in Fig.~\ref{fig_Diffrection_Sim}(d), which is the so-called SNS.
Moreover, as the size of the obstacle increases, the invisible region also expands, as indicated by the attenuation region from $S_d^{(1)} S_u^{(1)}$ ($\Delta S^{(1)} = 180\,\text{mm}$) to $S_d^{(2)}S_u^{(2)}$ ($\Delta S^{(2)} = 300\,\text{mm}$) in Fig.~\ref{fig_Diffrection_Sim}(d) and (e), respectively.
%
Meanwhile, some peaks remain observable \textit{both inside and outside} the shadowed region, as highlighted in Fig.~\ref{fig_Diffrection_Sim}(e). 
The peaks inside the highlighted shadow region responses are mainly caused by \textbf{edge diffraction}, which can be viewed as a special scattering phenomenon that occurs at knife edges or sharp wedge corners~\cite[Chapters 4-5]{bertoni2001radio}.

These observations suggest that \textbf{blockage is captured by partial visibility and diffraction is treated as an additional edge-induced inhomogeneous scattering component}.
Hence, the combined effects of blockage and diffraction give rise to mechanism-dependent spatial signatures, as further illustrated in Fig.~\ref{fig:difrraction_1}.

%
\subsection{Spherical Wavefront Parametric Channel Model}
Consider an ELAA system with  $M$ Tx and $N$ Rx antennas and employing a non-overlapping frequency-division multiplexing (FDM) waveform. The FDM frame consists of $P$ sub-bands, each with bandwidth $f_s$.
Let $\mathbf{z}\in\mathbb{C}^{MNP}$ denote the baseband-equivalent channel vector, where $m=1,2,\ldots,M$ and $n=1,2,\ldots,N$ index the Tx and Rx array elements, respectively, and $p=1,2,\ldots,P$ indexes the sub-band. 
The channel is modeled as the superposition of $L$ multipaths as
%
\begin{align}\label{eq:ch_1}
\mathbf{z} = \sum_{l=1}^{L} \mathbf{z}_l + \mathbf{w}
= \sum_{l=1}^{L} \alpha_{l}\!
\left[\left(\boldsymbol{\gamma}_{l} \odot \Delta \boldsymbol{\alpha}_{l}\right)\!\otimes\!\boldsymbol{1}\right]\!
\odot e^{\mathbf{a}_l(\mathbf{\tau})} + \mathbf{w},
\end{align}
where $l=1,2,\ldots,L$ indexes the multipath components, $\alpha_{l}$ denotes the stationary complex gain of the $l$th path while $\boldsymbol{\gamma}_{l}$ and $\Delta \boldsymbol{\alpha}_{l}$ characterize the SNS effects, and $\mathbf{w} \sim \mathcal{N}\left(\mathbf{0}, \sigma_0^2 \mathbf{I}\right)$ denotes zero-mean Gaussian distribution noise with power $\sigma_0^2$. 
In particular,  
$\boldsymbol{\gamma}_{l} \in \mathbb{C}^{MN} = [\gamma_{m,n,l}]_{MN}$ denotes the visibility vector\footnote{In most of the works, visibility is typically based on physical blockage \cite{10447918,10509715}, while this paper also includes visibility due to inhomogeneous scattering like specular reflection and edge diffraction.}, with the $l$th path w.r.t. the $m$th Tx-$n$th Rx channel as
\begin{equation} \label{eq:ch_1_3}
\gamma_{m,n,l} = 
\begin{cases}
1,   & \text{the $l$th path is visible } \\
0,   & \text{the $l$th path is invisible}
\end{cases}  
\end{equation}
and $\Delta \boldsymbol{\alpha}_{l} \in \mathbb{C}^{MN} = [\alpha_{m,n,l}]_{MN}$ denotes the variation in SNS amplitude across the array\footnote{Besides the propagation mechanisms induced inhomogeneous multipaths, the variation in amplitudes can be attributed to wide-band effects \cite{10179246}.}, and $\boldsymbol{1} \in \mathbb{C}^{P}$ is a one vector.
$\mathbf{a}_l(\tau) \in \mathbb{C}^{ MNP} $ denotes the steering vector determined by the absolute delay under the spherical wavefront assumption as
\begin{equation} \label{eq:ch_vector_1}
     \mathbf{a}_l(\mathbf{\tau})  =  
    \left[  {\tau_{m,n,l}} \right]_{MN} \otimes \left[ {-j2\pi f_p}\right]_{P},
\end{equation}
where $f_p = p f_s$ and $\tau_{m,n,l}$ denotes delay as
\begin{align} \label{eq:delay_0}
\tau_{m,n,l} =  \frac{d_{m,n,l}}{c},
\end{align}
where $c$ is the speed of light and $d_{m,n,l}$ is the exact propagation distance of the $l$th path from the $m$th Tx to the $n$th Rx antenna. 
\begin{rem}
    {\em Near-Field v.s. Far-Field:} 
    In far-field models, AoA and AoD are directly represented in the steering vector $\mathbf{a}_l(\tau)$ and are constant for each pair of Tx-Rx channels. 
    In near-field models, the range-dependent nature of spherical wavefronts is a constraint to regulate the delay-dependent angles in the steering vector $\mathbf{a}_l(\tau)$. 
    Equation~\eqref{eq:ch_vector_1} provides a unified representation that covers both near-field and far-field models. 
    When the propagation distance is sufficiently larger than the Rayleigh distance, the angle variation across the array will be negligible, and the corresponding steering vector naturally reduces to the conventional planar-wave approximation. 
\end{rem}
\begin{rem}
    {\em Geometric Interpretation of Path Delay:}
    The distance $d_{m,n,l}$ of each path is determined by the locations of the interaction points, i.e., scatterers and reflectors, along the propagation path. 
    As a result, $d_{m,n,l}$ carries explicit geometric information and forms the basis for constructing geometry-based constraints for scatterer and reflector localization under different propagation mechanisms.
\end{rem}

\subsection{Propagation Mechanisms and Geometry Constraints}

\subsubsection{Multi-Bounce Delay}
Using graph-based multi-bounce notation in \cite{Yuan_TWC25}, 
$d_{m,n,l}$ is calculated as
\begin{equation} \label{eq:delay_1_1}
\begin{aligned}
d_{m,n,l} =  
\begin{cases}
\| \mathbf{r}_{m,n,l} - \mathbf{r}_{\text{Tx},m} \|_2  +  
\| \mathbf{r}_{m,n,l} - \mathbf{r}_{\text{Rx},n} \|_2,  \text{one-bounce}, \\
\| \mathbf{r}_{m,n,l}^{1} - \mathbf{r}_{\text{Tx},m} \|_2  +  ... + \| \mathbf{r}_{m,n,l}^{K-1} - \mathbf{r}_{m,n,l}^{K} \|_2  \\   
   ~~~~~~~ + \| \mathbf{r}_{m,n,l}^{K} - \mathbf{r}_{\text{Rx},n} \|_2,  
 ~~~~~~ ~~~~~~~ \text{multi-bounce}.
\end{cases}  
\end{aligned}
\end{equation}
where $\mathbf{r}_{\text{Tx},m}$, $\mathbf{r}_{\text{Rx},n}$, $\mathbf{r}_{m,n,l}^k$ denotes the coordinates of the $m$th Tx element, $n$th Rx element, and the $k$th reflector/scatterer of the $l$th path w.r.t. this channel, respectively, with $K$ denotes the bouncing-order of that path and $k=1,\dots,K$.  
For a $K$-bounce path, the signal propagates sequentially through a set of $K$ scatterers, with coordinates $\{\mathbf{r}_{m,n,l}^{k}\}_K$; particularly, for a one-bounce path, we omit the bouncing order of that scatterer/reflector as $\mathbf{r}_{m,n,l}^1 \triangleq \mathbf{r}_{m,n,l}$. 

\subsubsection{Scattering and Specular Reflection}
%

Using $m_0 \in [1,M]$ and $n_0 \in [1,N]$ to denote the index of reference Tx and Rx, respectively\footnote{The reference channel can be any single-input single-output (SISO) channel of the multiple-input multiple-output (MIMO) systems, whereas, typically, $\text{Tx}_1$ and $\text{Rx}_1$  are chosen as reference antennas.}, the GC model of \textbf{scattering} paths in \eqref{eq:delay_1_1}, can be characterized as 
\begin{equation} \label{eq:gc_s1}
 \mathbf{r}_{m,n,l}^k = \mathbf{r}_{m_0,n_0,l}^k,   ~~  \gamma_{m,n,l} = \gamma_{m_0,n_0,l},
\end{equation}
where all different Tx-Rx pairs receive a spatially stationary path $l$ from a coherent scatterer.  

In contrast, the GC model of \textbf{reflection} paths in \eqref{eq:delay_1_1}, would be further characterized as
\begin{equation} \label{eq:GC_1}
\begin{cases}
 & \mathbf{r}_{m,n,l}^k   \neq \mathbf{r}_{m_0,n_0,l}^k,   ~~   \gamma_{m,n,l} \neq \gamma_{m_0,n_0,l}, \\
 & \mathbf{n}_{m,n,l}^k = \mathbf{n}_{m_0,n_0,l}^k,      
\end{cases}  
\end{equation}
where the specular illumination leads to SNS and element-wise variations across the array.
Nevertheless, for a smooth reflecting surface, although these bouncing points vary across antenna elements, the associated surface normal vector remains parallel as illustrated in Fig.~\ref{fig:paralell_normals}, 
\begin{equation}  \label{eq:ch_con_3_0}
    \mathbf{n}_{m_0,n_0,l}^k =   \dfrac{ \mathbf{r}_{m_0,n_0,l}^k - \mathbf{r}_{m_0,n_0,l}^{k-1} }{ \| \mathbf{r}_{m_0,n_0,l}^k - \mathbf{r}_{m_0,n_0,l}^{k-1} \|_2  } + 
        \dfrac{ \mathbf{r}_{m_0,n_0,l}^k - \mathbf{r}_{m_0,n_0,l}^{k+1} }{ \| \mathbf{r}_{m_0,n_0,l}^k - \mathbf{r}_{m_0,n_0,l}^{k+1} \|_2  }.
\end{equation}
Particularly in \eqref{eq:ch_con_3_0}, if $k=1$, $\mathbf{r}_{m,n,l}^0$ will be replaced by $\mathbf{r}_{\text{Tx},m}$; if $k=K$, $\mathbf{r}_{m,n,l}^{K+1}$ will be replaced by $\mathbf{r}_{\text{Rx},n}$. Therefore, the normal vector of one-bounce reflection path is   
\begin{equation} \label{eq:ch_con_3}
    \mathbf{n}_{m_0,n_0,l} =   \dfrac{ \mathbf{r}_{m_0,n_0,l} - \mathbf{r}_{\text{Tx},m_0} }{ \| \mathbf{r}_{m_0,n_0,l} - \mathbf{r}_{\text{Tx},m_0} \|_2  } + 
        \dfrac{ \mathbf{r}_{m_0,n_0,l} - \mathbf{r}_{\text{Rx},n_0} }{ \| \mathbf{r}_{m_0,n_0,l} - \mathbf{r}_{\text{Rx},n_0} \|_2  }.
\end{equation}



\subsubsection{Blockage and Diffraction}
The GC model of \textbf{diffraction} paths in \eqref{eq:delay_1_1} can be treated in our algorithms as  
\begin{equation} \label{eq:gc_dif}
 \mathbf{r}_{m,n,l}^k = \mathbf{r}_{m_0,n_0,l}^k,   ~~  \gamma_{m,n,l} \neq  \gamma_{m_0,n_0,l},
\end{equation}
where all different Tx-Rx pairs receive a coherent scattering source, while with different visibility to that source.  

\subsection{Problem formulation}
%
Given the measurement data $\mathbf{y} \in \mathbb{C}^{MNP}$, which is decomposed of $L$ hidden multipaths\footnote{Although the exact number of paths $L$ is unknown, the Akaike information criterion can estimate $L$ from a statistical perspective \cite{new_look_1974}. In practice, we may initially estimate a rough upper bound on the number of multipaths and then refine $L$ based on the convergence of the objective function \cite{liu2024debris}.} as
\begin{equation} 
\begin{aligned} \label{eq:prob_0}
    \mathbf{y} \triangleq \sum_{l=1}^{L} {\mathbf{z}_l}(\bm{\Theta}_l) + \beta_l \mathbf{w},
\end{aligned}
\end{equation}
where $\sum_{l=1}^{L} \beta_l^2 = 1$ is used to constrain the noise power, and $\bm{\Theta}_l = [\bm{\theta}_{m,n,l}]_{MN}$ represents the channel parameters of the $l$th multipath, with entry 
\begin{align}
& \bm{\theta}_{m,n,l} \triangleq  \begin{cases}
\big[\alpha_l, \Delta \alpha_{m,n,l}, \gamma_{m,n,l}, \mathbf{r}_{m,n,l}\big], & \text{one-bounce}, \\
\big[\alpha_l, \Delta \alpha_{m,n,l}, \gamma_{m,n,l}, \{\mathbf{r}_{m,n,l}^k\}_{k=1}^2 \big], & \text{two-bounce}, \\
\big[\alpha_l, \Delta \alpha_{m,n,l}, \gamma_{m,n,l}, \tau_{m,n,l}\big], & \text{high-bounce}.
\end{cases} \label{eq:parameter}
\end{align}
The problem is formulated as  
\begin{subequations}
\begin{align}
& \mathcal{P}_1 \quad \arg \min_{\bm{\Theta}_l} \left\| \mathbf{y} - \sum_{l=1}^{L} \mathbf{z}_l(\bm{\Theta}_l) \right\|_2, \label{eq:prob_1} \\
& \text{s.t.} \quad \mathbf{r}_{m,n,l}^k  = \mathbf{r}_{m_0,n_0,l}, \quad \text{scatterers}, \label{eq:ch_con_1_1} \\
& \quad \quad \mathbf{n}_{m,n,l}   = \mathbf{n}_{m_0,n_0,l}, \quad \text{reflectors}. \label{eq:ch_con_2}
\end{align}
\end{subequations}
where, as noted in \cite{Yuan_TWC25}, identifying the exact coordinates for high-bounce paths (i.e., paths involving more than two bounces) can lead to ambiguity. Therefore, in \eqref{eq:parameter}, we estimate only the delay for these high-bounce paths. This preserves the E-M iteration, while for one-bounce and two-bounce paths, we can recover the exact scatterer locations.

\section{GC-SAGE-based Localization and Mapping}\label{Sec:Algo}
%
The problem $\mathcal{P}_1$ is a non-convex multi-object multivariate problem, so it is natural to follow the iterative E-M style of the SAGE~\cite{fleury1999channel}. 
The E-step remains similar to conventional SAGE, while the key difference lies in the M-step, where different GCs are applied in joint parameter estimation and mapping.
We first introduce the scattering-dominated channel estimation in Section~\ref{sec:multi-bounce-scatter}, which can deal with scattering multi-bounce paths and partial blockage, forming the baseline for the GC-SAGE algorithm. 
We then extend the framework to the more general hybrid reflection-scattering channel.
%
\subsection{Multi-Bounce Scattering Paths with SNS}\label{sec:multi-bounce-scatter}
%
We first estimate the $L$ path parameters in the chosen reference channel, namely the amplitudes $\alpha_{l}$ and delays $\tau_{m_0,n_0,l}$. Subsequently, the \textbf{E-step} and \textbf{M-step} are applied iteratively to estimate the spatial parameters of each path by leveraging the full MIMO channel.
In the $i$th iteration, for $l = 1, 2, ..., L$, 
\begin{equation} \label{eq:EM}
\begin{aligned}
\begin{cases}
 \text{E-step:}~ \hat{\mathbf{y}}_l^{(i)}  = 
    \mathbf{z}_l( \hat{\bm{\Theta}}_{l}^{(i-1)} ) + \beta_l \left( \mathbf{y} - \sum_{l=1}^{L} \mathbf{z}_l( \hat{\bm{\Theta}}_{l}^{(i-1)} ) \right),     \\
    \text{M-step:}~ \hat{\bm{\Theta}}_{l}^{(i)} =  \arg \underset{ \hat{\bm{\Theta}}_{l} }{\min} \dfrac{\left( \hat{\mathbf{y}}_l^{(i)} - \mathbf{z}_l(\mathbf{\Theta}_{l} ) \right)^{H}  \left( \hat{\mathbf{y}}_l^{(i)} - \mathbf{z}_l(\mathbf{\Theta}_{l} ) \right) }{\beta_{l} \sigma_0^2},
\end{cases}  
\end{aligned}
\end{equation}
where $\mathbf{\hat{y}}_l^{(i-1)}$ is the estimated signal of the $l$th path based on parameters of the results of the $(i-1)$th M-step.
Initially, for $i=1$, $\tau_{m,n,l}^{(0)}$ is replaced by $\tau_{m_0,n_0,l}$ to start the iteration. 
For $i > 1$, they phase-related parameter $\tau_{m,n,l}^{(i)} = \dfrac{d_{m,n,l}^{i}}{c}$ is calculated by localized coordinates as \eqref{eq:delay_1_1}.  

\subsubsection{Localization of Scatterers}
To locate scatterers of the $l$th path in the $i$th iteration, we adopt both GCs on the distance-domain \eqref{eq:delay_1_1} and spatial domain \eqref{eq:gc_s1}. 

We first assume the $l$th path is a \textbf{one-bounce} path, i.e., the propagation track $\mathrm{Tx}\rightarrow \mathrm{scatter}\rightarrow \mathrm{Rx}$. 
With the known Tx and Rx coordinates, the candidate scatterer locations can be constrained to an ellipsoidal surface\footnote{In the bistatic sensing model, for $d_{m,n,l} > \| \mathbf{r}_{\mathrm{Tx},m} - \mathbf{r}_{\mathrm{Rx},n} \|$, the interaction point $\mathbf{r}_{m,n,l}$ lies on an ellipsoidal surface in 3D space whose foci are $\mathbf{r}_{\mathrm{Tx},m}$ and $\mathbf{r}_{\mathrm{Rx},n}$ \cite{8642926}.} as
\begin{equation}
\begin{aligned}\label{eq:ellipsoidal_surface}
& \mathcal{E}_{\mathrm{ell}}\!\left(d_{m_0,n_0,l}\right)
 \triangleq  \\
& \left\{
\mathbf{r}\in \mathcal{V}:
\left\|\mathbf{r}-\mathbf{r}_{\mathrm{Tx},m_0}\right\|_2 + \left\|\mathbf{r}-\mathbf{r}_{\mathrm{Rx},n_0}\right\|_2
= d_{m_0,n_0,l} \right\},
\end{aligned}
\end{equation}
where $\mathcal{V} \subset \mathbb{R}^3$ is the defined $3$-D search space\footnote{The searching space $\mathcal{V}$ can be defined as the bounded sensing region, for example, a room or a venue of known dimensions. While the overall size of $\mathcal{V}$ is assumed known, its detailed geometry and the objects are to be estimated.}.
%
Then, locate the one-bounce scatterer as
\begin{subequations}\label{eq:m_step_one_bounce}
\begin{align} 
& \hat{\mathbf{r}}_{m,n,l}^{(i)}  = 
 \arg \underset{ \hat{\mathbf{r}}_{m,n,l}^{(i)} }{\min} \dfrac{ 1  }{\beta_{l} \sigma_0^2} \left\| \hat{\mathbf{y}}_l^{(i)} - \mathbf{z}_l( \mathbf{r}_{m,n,l} ) \right\|_2^{2}  , \\
& \text{s.t.} ~~~~~ \mathbf{r}_{m_0,n_0,l} \in \mathcal{E}_{\mathrm{ell}}\!\left(d_{m_0,n_0,l}\right), \\
& ~~~~~~ \quad \mathbf{r}_{m,n,l}^k   \triangleq \mathbf{r}_{m_0,n_0,l}.
\end{align}
\end{subequations}

Then we assume the $l$th path to be \textbf{two-bounce}, i.e., the propagation track $\mathrm{Tx}\rightarrow \mathrm{scatter~1} \rightarrow \mathrm{scatter~2} \rightarrow \mathrm{Rx}$. 
For the two-bounce path, the corresponding delay constraint can be decomposed into a nested two-loop search. Specifically, we fix the first scatterer location $\mathbf{r}_{m_0,n_0,l}^1$ to perform a one-bounce-like search, and iterate over all candidates $\mathbf{r}_{m_0,n_0,l}^1 \in \mathcal{V}$.
\begin{equation}
\begin{aligned}\label{eq:ellipsoidal_surface_2}
& \mathcal{E}_{\mathrm{ell}}\!\left(d_{m_0,n_0,l}, \mathbf{r}_{m_0,n_0,l}^1 \right)
 \triangleq \{ \mathbf{r}\in \mathcal{V}:    \left\| \mathbf{r} - \mathbf{r}_{m_0,n_0,l}^1 \right\|_2 + \\
&  ~~
\left \|\mathbf{r}-\mathbf{r}_{\mathrm{Rx},n_0}\right\|_2 
= d_{m_0,n_0,l} - \left\| \mathbf{r}_{\text{Rx}_{n_0}} - \mathbf{r}_{m_0,n_0,l}^1 \right\|_2 \}.
\end{aligned}
\end{equation}
The localization step is similar to \eqref{eq:m_step_one_bounce}.

\textbf{Bounce Order}: 
The parameters are initially set for high-bounce estimation using conventional SAGE \cite{feder1988parameter}. The bounce order is determined by selecting the configuration that minimizes the objective function in \eqref{eq:prob_1}. 
If neither the one-bounce nor two-bounce search minimizes the objective, we classify the $l$th path as a high-bounce path, where we retain the delay parameters, allowing the iterations to proceed. The updates terminate once the objective function converges.

\subsubsection{SNS Detection}: 
Directly estimating the sparse parameter $\boldsymbol{\gamma}_{l}$ and the attenuation parameter $\Delta \boldsymbol{\alpha}_{l}$ under blockage and diffraction is computationally expensive. For example, based on the quantities in \eqref{eq:ch_1}, the additional computational complexity scales as $\mathcal{O}\!\left(L2^{MN}\right)$.
A practical way is to estimate the equivalent amplitude as
\begin{equation}
    \begin{aligned} \label{eq:mstep_3}
        \tilde{\mathbf{\alpha}}_{m,n,l}^{(i)} & \triangleq  \alpha_{l}^{(i)} \mathbf{\gamma}_{m,n,l}^{(i)} \Delta \mathbf{\alpha}_{m,n,l}^{(i)} \\
         & = \left( \mathbf{z}_{l}([ \alpha_l, \Delta\mathbf{\alpha}_l, \mathbf{\gamma}_{l}, \hat{\tau}_{m,n,l}^{(i)} ] )^H \mathbf{z}_{l}([ \alpha_l, \Delta\mathbf{\alpha}_l, \mathbf{\gamma}_{l}, \hat{\tau}_{m,n,l}^{(i)} ] ) \right)^{-1} \\ 
        & ~~ \times \mathbf{z}_{l}([  \alpha_l, \Delta\mathbf{\alpha}_l, \mathbf{\gamma}_{l}, \hat{\tau}_{m,n,l}^{(i)} ] )^H \hat{\mathbf{y}}_{l}^{(i)}.
    \end{aligned}
\end{equation}
\begin{rem}
{\em} 
In blocked paths, the estimated amplitudes are typically very weak or at the noise floor, e.g., results in Fig.~\ref{fig:simu_case_1}(d). 
These amplitude estimates implicitly capture both path existence and effective propagation.
Therefore, the SNS effects are accounted for, and its parameters do not affect the E-step or M-step when estimating other paths.
\end{rem}
\begin{algorithm}[t]
\caption{GC-SAGE under Scattering-Only Assumption}\label{algo:DM_SAGE}
\begin{algorithmic}[1] 
    \State \textbf{Input:} $\mathbf{y}$, $\mathbf{z}$, $L$ 
    \State \textbf{Output:} $\bm{\Theta} = [ \bm{\Theta}_l ]_L$
    \State \textbf{Initialization:} Using conventional SAGE \\
        Estimate $\{ \alpha_{0,l}, \tau_{0,l} \}_L$ of reference channel
    \For{ $i$th iteration, $i \geq 1$}
        \For{$l$ = 1, 2, ..., $L$}
        \State \textbf{E-step} using \eqref{eq:EM}
        \State \textbf{M-step} using \eqref{eq:ellipsoidal_surface} and \eqref{eq:m_step_one_bounce}
            \If{Obj \eqref{eq:prob_1} function decreased }
                \State Estimate $\bm{\Theta}_l$ as one-bounce set in \eqref{eq:parameter}
            \Else
                \State \textbf{M-step} using \eqref{eq:m_step_one_bounce} and \eqref{eq:ellipsoidal_surface_2} 
                \If{ Obj \eqref{eq:prob_1} function decreased  }
                    \State Estimate $\bm{\Theta}_l$ as two-bounce set in \eqref{eq:parameter}
                \Else
                    \State \textbf{M-step} in \eqref{eq:EM}: as high-bounce paths  
                    \State Estimate $\bm{\Theta}_l$ as high-bounce set in \eqref{eq:parameter}
                \EndIf
            \EndIf
        \EndFor
    \EndFor
\end{algorithmic}
\end{algorithm}
The overall implementation of scattering-only assumption GC-SAGE is summarized in Algorithm~\ref{algo:DM_SAGE}. 


%
%
\begin{figure}[t]
     \centering
    \includegraphics[width=0.45\textwidth]{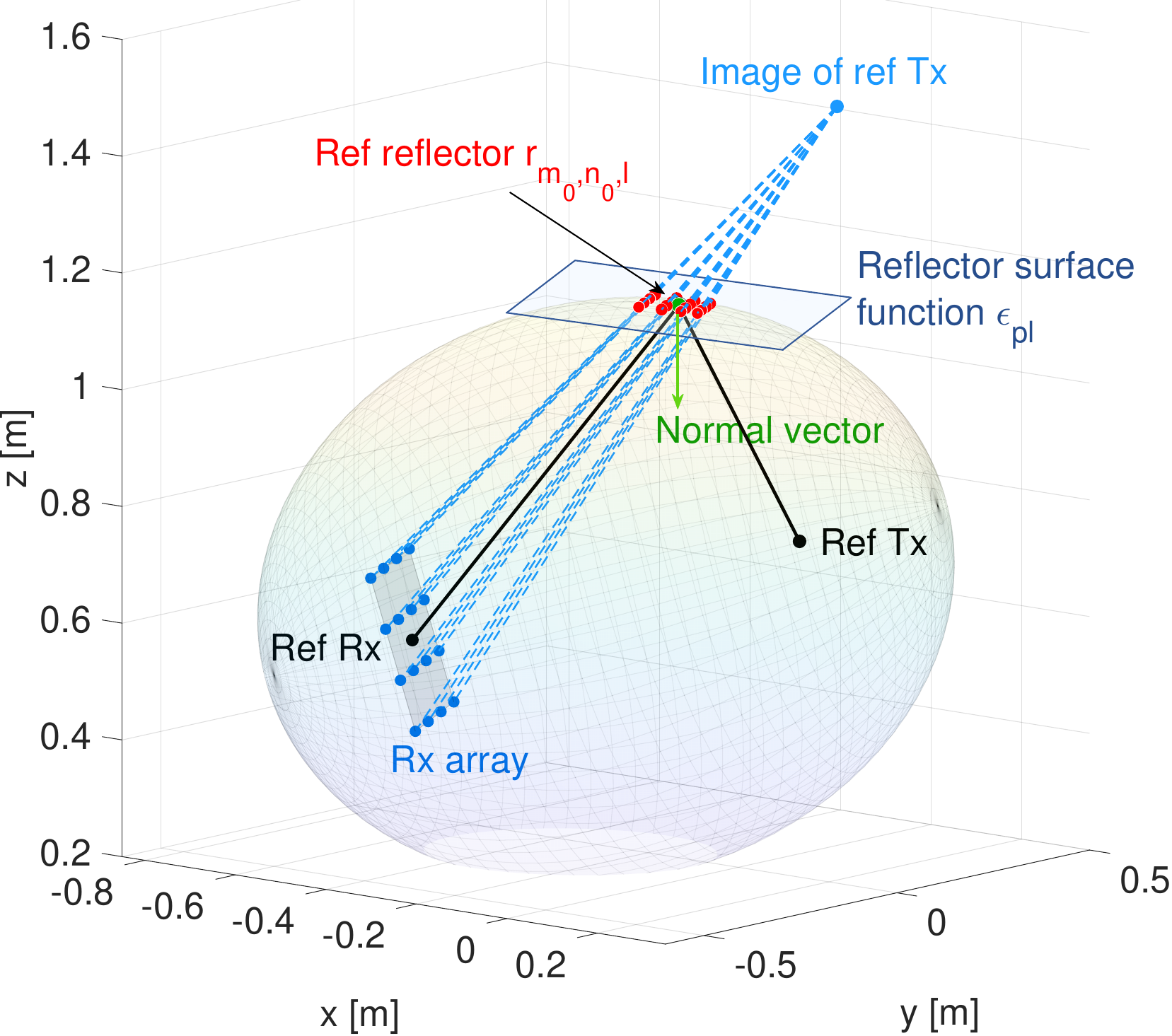}
        \caption{Illustration of the scatterers and reflectors localization in M-step based on geometry constraints. 
        }
  \label{fig:search_app}
\end{figure}
\begin{algorithm}[t]
\caption{Reflector localization in M-step}\label{algo:GC_SAGE}
\begin{algorithmic}[1] 
    \State \textbf{Input:} $\mathbf{y}$, $\mathbf{z}$, $L$ 
    \State \textbf{Output:} $\bm{\Theta} = [ \bm{\Theta}_l ]_L$
    \State \textbf{Initialization:} M-step in Algorithm~\ref{algo:DM_SAGE}, calculate the value of objective \eqref{eq:prob_1} $\text{val}_{l,s}$ 
            \State Apply \eqref{eq:m_step_one_bounce_reflection}, locate $\left[ \hat{\mathbf{r}}_{m,n,l} \right]_{MN}$ ~~(Reflector case)
            \State Calculate the value of objective \eqref{eq:prob_1} $\text{val}_{l,r}$
            \If{$\text{val}_{l,r} < \text{val}_{l,s} $ \textbf{and} $\text{val}_{l,r} \leq \text{val}_{l}^{(i-1)}$ } 
            \State Update $\left[ \hat{\mathbf{r}}_{m,n,l} \right]_{MN}$ in $\hat{\bm{\theta}}_l^{(i)} $, $\text{val}_{l}^{(i)} = \text{val}_{l,r} $
            \Else
            \EndIf
\end{algorithmic}
\end{algorithm}
\vspace{-0.5cm}
%
\subsection{Hybrid Reflection-Scattering Propagation}
By including both reflection and scattering paths estimation, the \textbf{M-Step} of Algorithm~\ref{algo:DM_SAGE} requires additional reflection search. 
Here we take the $i$th iteration of the $l$th path, assuming that it was a one-bounce path, as an example. 
Based on the reference distance $d_{m_0,n_0,l}$ and coordinate of reference Tx and Rx, we could define the ellipsoidal surface $\mathcal{E}_{\mathrm{ell}}\!\left(d_{m_0,n_0,l}\right)$ by \eqref{eq:ellipsoidal_surface}.
We first treat it as a \textbf{scattering path}. Following \eqref{eq:m_step_one_bounce}, we iterate over the candidate points $\mathbf{r}_{m_0,n_0,l}$ on the ellipsoidal surface and, for each candidate, minimize \eqref{eq:m_step_one_bounce}. This yields the most likely scatterer location and the corresponding minimum objective value, denoted by $\text{val}_{l,s}$. 
%
We then treat it as a \textbf{reflection path}, where Fig.~\ref{fig:search_app} helps to illustrate the process. 
For each reflector candidate $\mathbf{r}_{m_0,n_0,l}$ on an ellipsoidal surface $\mathcal{E}_{\mathrm{ell}}\!\left(d_{m_0,n_0,l}\right)$, we can use \eqref{eq:ch_con_3} to calculate surface normal $\mathbf{n}_{m_0,n_0,l}$, illustrated as green arrow in Fig.~\ref{fig:search_app}, and the corresponding reflection plane $\mathcal{E}_{\mathrm{pl}} \left(\mathbf{r}_{m_0,n_0,l},\mathbf{n}_{m_0,n_0,l}\right)$ as 
\begin{equation}
\begin{aligned}\label{eq:mstep_1}
\mathcal{E}_{\mathrm{pl}} & \left(\mathbf{r}_{m_0,n_0,l},\mathbf{n}_{m_0,n_0,l}\right)
 \triangleq \left\{
\mathbf{r}\in\mathcal{V}: 
\mathbf{n}_{m_0,n_0,l}^{T}\!\left(\mathbf{r}-\mathbf{r}_{m_0,n_0,l}\right)=0
\right\}.
\end{aligned}
\end{equation}
Then, calculate the mirror image $\mathbf{r}_{\text{Tx},m_0}'$ of the reference Tx because of $\mathcal{E}_{\mathrm{pl}} \left(\mathbf{r}_{m_0,n_0,l},\mathbf{n}_{m_0,n_0,l}\right)$ as
\begin{equation}\label{eq:mirror_tx}
\mathbf{r}_{\mathrm{Tx},m_0}' 
= \mathbf{r}_{\mathrm{Tx},m_0} -2\,\mathbf{n}_{m_0,n_0,l}\, \mathbf{n}_{m_0,n_0,l}^{T} \left(\mathbf{r}_{\mathrm{Tx},m_0}-\mathbf{r}_{m_0,n_0,l}\right).
\end{equation}
The set of all straight lines from $ \mathbf{r}_{\text{Tx},m_0}' $ to each $ \mathbf{r}_{\text{Rx},n}$ is 
\begin{equation}\label{eq:lines_image}
\mathcal{L}\!\left( \mathbf{r}_{\text{Tx},m_0}', \left[\mathbf{r}_{\text{Rx}_n}\right]_N \right)\!\triangleq\!\left\{ \mathbf{r} \in \mathcal{V}\!:\! 
\frac{\mathbf{r} - \mathbf{r}_{\text{Tx},m_0}'}{\|\mathbf{r} - \mathbf{r}_{\text{Tx},m_0}'\|_2}\!=\!\frac{\mathbf{r} - \mathbf{r}_{\text{Rx},n}}{\|\mathbf{r} - \mathbf{r}_{\text{Rx},n}\|_2}
\right\}.
\end{equation}
The intersection between the lines $\mathcal{L}\!\left( \mathbf{r}_{\text{Tx},m_0}', \left[\mathbf{r}_{\text{Rx}_n}\right]_N \right)$ and plane $\mathcal{E}_{\mathrm{pl}} \left(\mathbf{r}_{m_0,n_0,l},\mathbf{n}_{m_0,n_0,l}\right)$ are reflecting point specific to $\text{Tx}_{m_0}$ to N Rx channels, i.e., 
\begin{equation}\label{eq:mstep_2}
\begin{aligned}
\mathbf{r}_{m_0,n,l} =
\mathcal{L}\!\left( \mathbf{r}_{\text{Tx},m_0}', \left[\mathbf{r}_{\text{Rx}_n}\right]_N \right) \cap \mathcal{E}_{\mathrm{pl}} \left(\mathbf{r}_{m_0,n_0,l},\mathbf{n}_{m_0,n_0,l}\right) .
\end{aligned}
\end{equation}
Iterating on all Tx antennas as a reference, the set of reflection candidates $\{\mathbf{r}_{m,n,l}\}_{M \times N}$ based on $d_{m_0,n_0,l}$ will be obtained, which are shown as the red dots\footnote{Note the red dots are not necessarily on the ellipsoid, while $\mathbf{r}_{m_0,n_0,l}$ is constrained to $\mathcal{E}_{\mathrm{pl}}\!\left(\mathbf{r}_{m_0,n_0,l},\mathbf{n}_{m_0,n_0,l}\right)$.
The reflection plane is defined as the tangent plane to $\mathcal{E}_{\mathrm{pl}}\left(\mathbf{r}_{m_0,n_0,l},\mathbf{n}_{m_0,n_0,l}\right)$ at
$\mathbf{r}_{m_0,n_0,l}$.} in Fig.~\ref{fig:search_app}. 
%

Each reference point $\mathbf{r}_{m_0,n_0,l}$ on the ellipse corresponds to a set of candidate reflection planes, as defined in \eqref{eq:mstep_2}. These candidate sets are selected by minimizing the following objective function as
\begin{subequations}\label{eq:m_step_one_bounce_reflection}
\begin{align} 
& \hat{\mathbf{r}}_{m,n,l}^{(i)}  = \arg \underset{ \hat{\mathbf{r}}_{m,n,l}^{(i)} }{\min} \dfrac{1}{\beta_{l} \sigma_0^2}\left\| \hat{\mathbf{y}}_l^{(i)} - \mathbf{z}_l( \mathbf{r}_{m,n,l} ) \right\|_2^2 , \\
& \text{s.t.}~ \mathbf{r}_{m,n,l}\!=\!\mathcal{E}_{\mathrm{pl}}\left(r_{m_0,n_0,l}, \mathbf{n}_{m_0,n_0,l} \right) \cap \mathcal{L}\!\left( \mathbf{r}_{\text{Tx},m}', \mathbf{r}_{\text{Rx}_n} \right),\!
\end{align}
\end{subequations}
where the most likely reflectors are obtained and the corresponding minimized value $\text{val}_{l,r}$. The detailed operation of reflector localization is described in \textbf{Algorithm~\ref{algo:GC_SAGE}}.




\subsection{Complexity Analysis}\label{sec:complexity}
%
Let $V = \| \mathcal{V} \|_0$ denote the search space. 
In the \textbf{scattering-only} model in Algorithm~\ref{algo:DM_SAGE}, the computational complexity of a one-bounce brute-force search is $\mathcal{O}(LV)$. By constraining the coordinates to the elliptical trajectory, the complexity reduces to $\mathcal{O}(\eta LV)$, where $\eta \approx 2.29\%$ in simulation examples \cite{Yuan_TWC25}. The computational complexity of a two-bounce brute-force search is $\mathcal{O}(\eta^2 LV^2)$. The total computational complexity is therefore $\mathcal{O}(MN \times (\eta LV + \eta^2 LV^2))$.
By considering the \textbf{hybrid reflection-scattering} model in Algorithm~\ref{algo:GC_SAGE}, each path requires an additional reflection search. Although the reflectors are element-wise, they require an additional search using GCs before proceeding to the cost function in \eqref{eq:m_step_one_bounce_reflection}. The computational complexity minimization function is similar to scattering, although the element-wise reflecting points of each channel vary. 
The computational complexity for reflector localization is $\mathcal{O}(MN \times (\eta LV + \eta^2 LV^2))$.
In total, the complexity of the GC-SAGE algorithm is $\mathcal{O}\left( \left( 2MN \left(\eta LV + \eta^2 LV^2\right) \right)  \right)$.


\section{Validation}\label{sec:simu}
In this section, we validate the proposed GC-SAGE for near-field ELAA channel estimation and environment mapping using both RT-simulation and field-measurement data. 
In the first two RT-based simulations, we can flexibly configure the scenario, antenna configurations, and evaluate specific propagation mechanisms. Besides, RT provides visualization of the propagation trajectory of each path in the digital map \cite{10559769}, enabling direct qualitative comparisons between the localized scatterers/reflectors and the ground truth.
In the real-world field measurements in a basement environment, it contains comprehensive propagation mechanisms and enables assessing the algorithm's robustness under practical hardware and environmental conditions.

\begin{table}[t]
    \caption{Configurations used in simulation case~1 and case~2 }
    \centering
    \begin{tabular}{lclclclc|c|c}
        \toprule
        Configurations &   Simulation Case~$1$  &   Simulation Case~$2$ \\
        \midrule
        Central frequency $f_c$ [GHz] & $30$  & $30$  \\
        Bandwidth $B$ [GHz] & $1$ & $1$  \\
        Sub-bandwidth $f_s$ [MHz] & $10$  & $10$  \\
        Number of sub-bands $P$ & $101$  & $101$ \\
        SNR [dB] & $20$ & $5-20$ \\  
        Grid size [m] & 0.1/0.2  & 0.1   \\
        Room space [m$^2$] & [${6.5 \times 6.5}$] & [${6.5 \times 6.5}$] \\
        NO. of Tx  $\times$ Rx    & $16 \times 121$ MIMO & $3 \times 200$ MIMO \\
        Reference Tx and Rx  & $(2.1, 4.1)$ $(3.3, 1)$  & $(1, 0)$ $(2, 0)$\\
        Antenna spacing & $ 0.5 \lambda $  & $ 0.75 \lambda $ \\
        \bottomrule
    \end{tabular}
    \label{tab:simu_configs_1}
     \vspace{-0.2cm}
\end{table}
\begin{figure*}[ht]
     \centering
          \subfigure[Layout of the simulation scenario and propagation tracks ]{\includegraphics[width=0.45\textwidth]{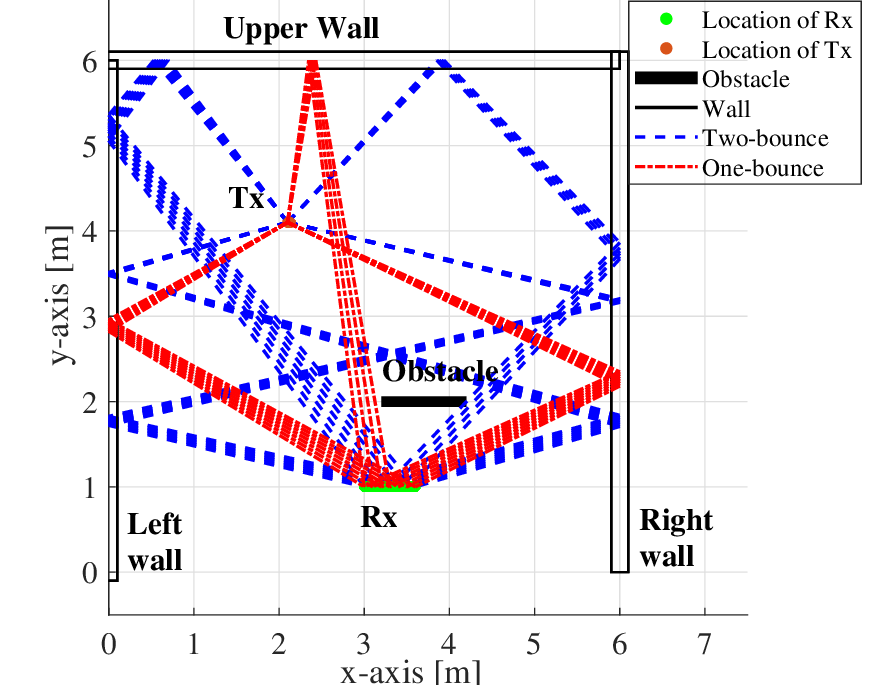}}
                        \hfill
    \subfigure[Concatenated power delay profile from $\text{Tx}_1$ to all Rxs ]{\includegraphics[width=0.45\textwidth]{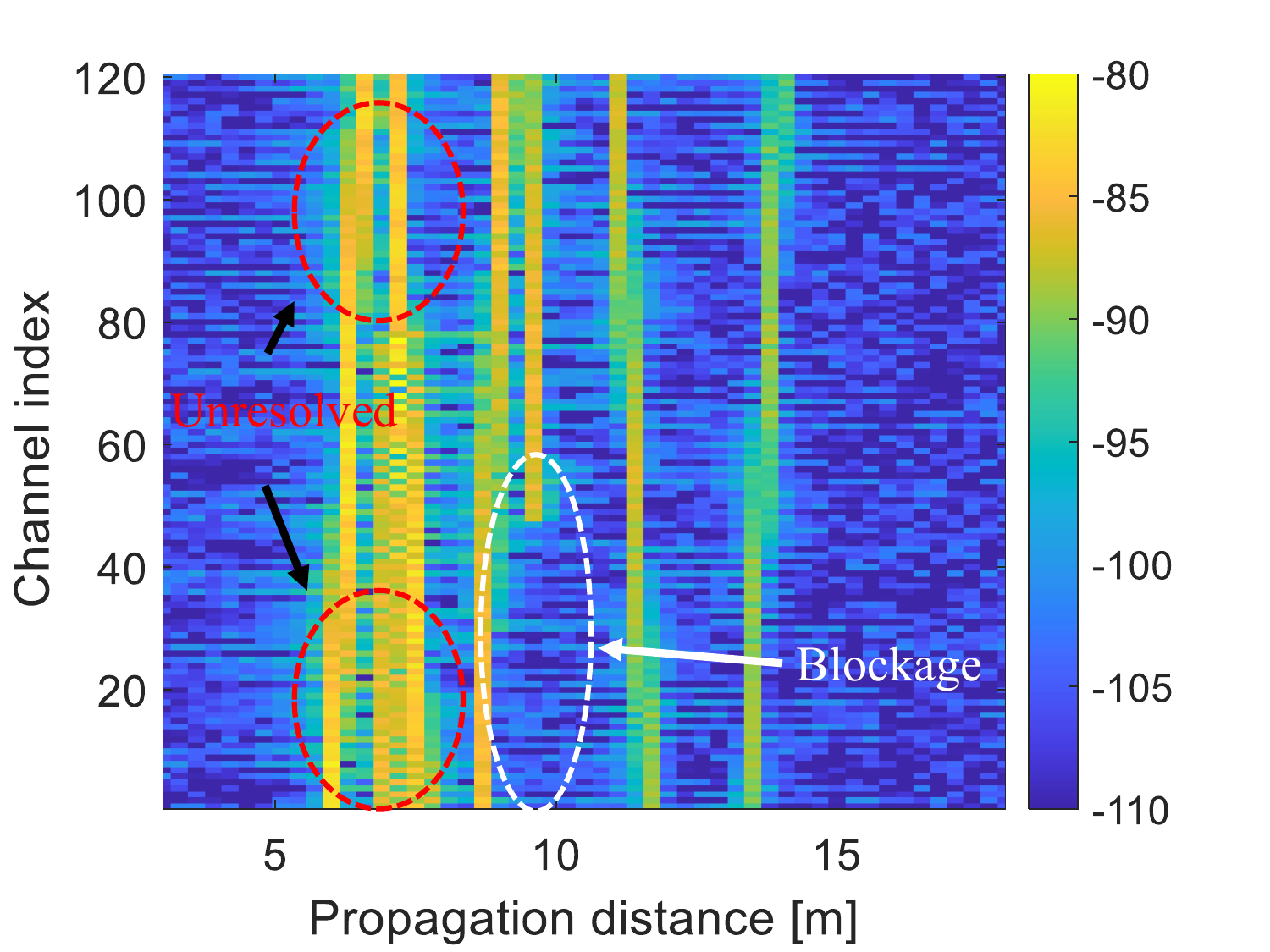}}
            \hfill
            \vspace{-0.2cm}
            \subfigure[Estimated scatterers and mapping results]{\includegraphics[width=0.45\textwidth]{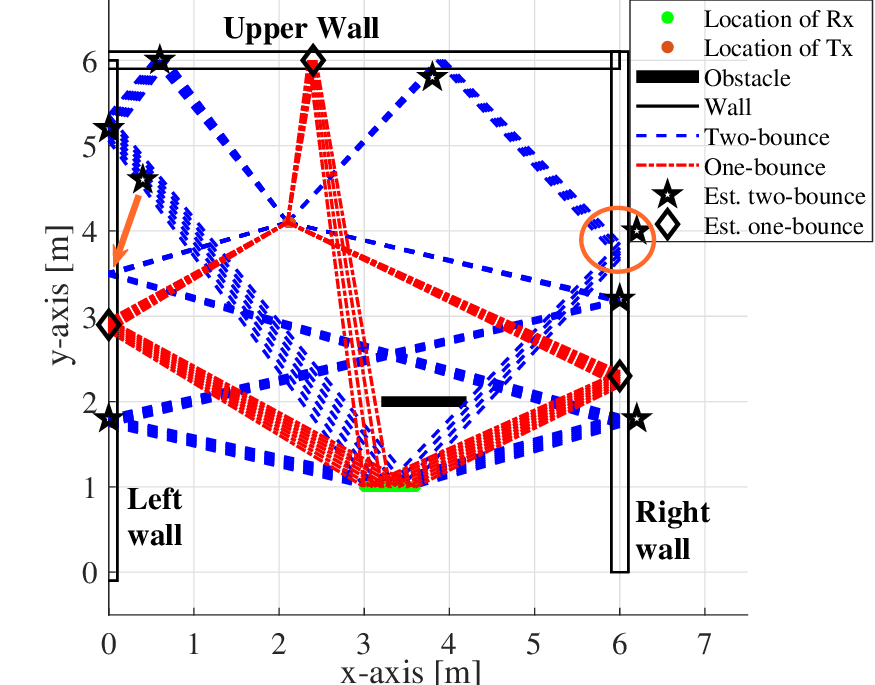}}
            \hfill
        \subfigure[Estimated amplitudes of all paths in the $\textbf{Tx}_1$ to Rx channels ]{\includegraphics[width=0.45\textwidth]{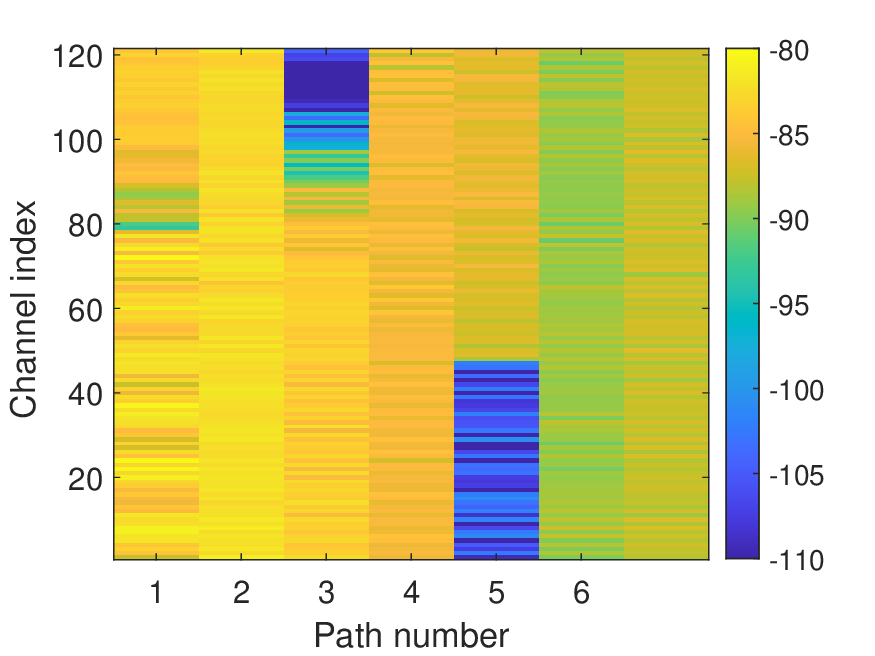}}
                        \hfill
                         \vspace{-0.2cm}
        \caption{Simulation case~1: (a) scenario layout and RT-generated propagation tracks; (b) CPDP across the receive aperture; (c) scatterer localization results; and (d) environment mapping results.}
    \label{fig:simu_case_1}
     \vspace{-0.5cm}
\end{figure*}
\subsection{Simulation Case~1: Multi-Bounce with Partial Blockage}
\subsubsection{Scenario}
We first validate the proposed algorithm in a simulated two-dimensional $2$-D SNS channel under a partial blockage scenario using a bistatic MIMO ELAA system, as shown in Fig.~\ref{fig:simu_case_1}(a).
The environment consists of three walls (left, upper, and right) and a single obstacle that partially blocks the receive aperture.
Unless otherwise stated, the simulation parameters follow Table~\ref{tab:simu_configs_1}.
%
The RT-generated propagation tracks for the one-bounce and two-bounce components are also illustrated in Fig.~\ref{fig:simu_case_1}(a).
Due to the obstacle, the two tracks, $\mathrm{Tx}\!\rightarrow\!\mathrm{upper~wall}\!\rightarrow\!\mathrm{Rx}$ and $\mathrm{Tx}\!\rightarrow\!\mathrm{upper~wall}\!\rightarrow\!\mathrm{right~wall}\!\rightarrow\!\mathrm{Rx}$, are partially blocked, resulting in a clear null region in the concatenated power delay profile (CPDP), as shown in Fig.~\ref{fig:simu_case_1}(b).
In addition, noticeable SNS effects beyond blockage are observed, such as range cell migration and unresolved multipath components.

\subsubsection{Estimation and Environment Reconstruction}
Fig.~\ref{fig:simu_case_1}(c) shows the reconstructed scatterers in the environment, where diamonds denote the localized one-bounce scatterers and stars denote the localized two-bounce scatterers.
The algorithm effectively separates the one-bounce and two-bounce paths, with relatively good accuracy in mapping the interaction points to the propagation track. 
There are two outlier two-bounce paths. The first scatterer of the path $\text{Tx} \rightarrow \text{left wall} \rightarrow \text{right wall} \rightarrow \text{Rx}$, highlighted by the orange arrow, is primarily due to the limited Tx aperture, which results in an insufficient Rayleigh distance to cover the entire region, introducing direction ambiguity.
Due to the {\em partial blockage} of Rx in the path $\text{Tx} \rightarrow \text{upper wall} \rightarrow \text{right wall} \rightarrow \text{Rx}$, the {\em reduced effective Rx array} affects both the Rayleigh distance and the signal strength, thus introducing ambiguity.
%
Table~\ref{tab:error_scatter_1} shows the quantified errors in scatterer localization.  
%
\begin{table}[t]
\caption{Quantified localization error of Simulation Case I}
\centering
\scriptsize
\setlength{\tabcolsep}{2.5pt}
\renewcommand{\arraystretch}{1.05}
\begin{tabularx}{\columnwidth}{p{1.05cm} p{2cm} X X p{1.05cm}}
\toprule
\makecell[l]{Bounce\\ order} & \makecell[l]{Path} & Truth & \makecell[l]{Estimation} & \makecell[l]{Error~[m]} \\
\midrule
\multirow{3}{*}{\makecell[l]{One-\\bounce}} & Left wall & $(0,\,2.88)$ & $(0,\,2.9)$ & $0.02$ \\
& \textit{Upper wall} & $(2.46,\,6)$ & $(2.4,\,6)$ & $0.06$ \\
& Right wall & $(6,\,2.27)$ & $(6,\,2.3)$ & $0.03$ \\
\midrule
\multirow{8}{*}{\makecell[l]{Two-\\bounce}} & \multirow{2}{*}{Upper and left wall}
& $(0.64,\,6)$ & $(0.6,\,6)$ & $0.04$ \\
&  & $(0,\,5.19)$ & $(0,\,5.2)$ & $0.01$ \\
& \multirow{2}{*}{\textit{Upper and right wall}} & $(3.94,\,6)$ & $(3.8,\,5.8)$ & $0.24$ \\
&  & $(6,\,3.84)$ & $(6.2,\,4)$ & $0.26$ \\
& \multirow{2}{*}{Left and right wall} & $(0,\,3.49)$ & $(0.4,\,4.6)$ & $1.18$ \\
&  & $(6,\,1.78)$ & $(6.2,\,1.8)$ & $0.10$ \\
& \multirow{2}{*}{Right and left wall} & $(6,\,3.19)$ & $(6,\,3.2)$ & $0.01$ \\
&  & $(0,\,1.78)$ & $(0,\,1.8)$ & $0.02$ \\
\bottomrule
\end{tabularx}
\label{tab:error_scatter_1}
 \vspace{-0.4cm}
\end{table}
\subsubsection{Blockage Paths Estimation and SNS Analysis}
Using the joint calculation in \eqref{eq:mstep_3}, the amplitude of each path in each channel, $\tilde{\mathbf{\alpha}}_{m,n,l}$, is calculated, and the SNS of each path is practically obtained, as shown in Fig.~\ref{fig:simu_case_1}(d).
The two blocked paths are captured in the scatterer mapping results in Fig.~\ref{fig:simu_case_1}(c). 
These findings demonstrate the robustness of the proposed method in partial blockage scenarios.

\begin{figure}[ht]
    \centering
    \subfigure[Benchmark: scattering-only model]{\includegraphics[width=0.42\textwidth]{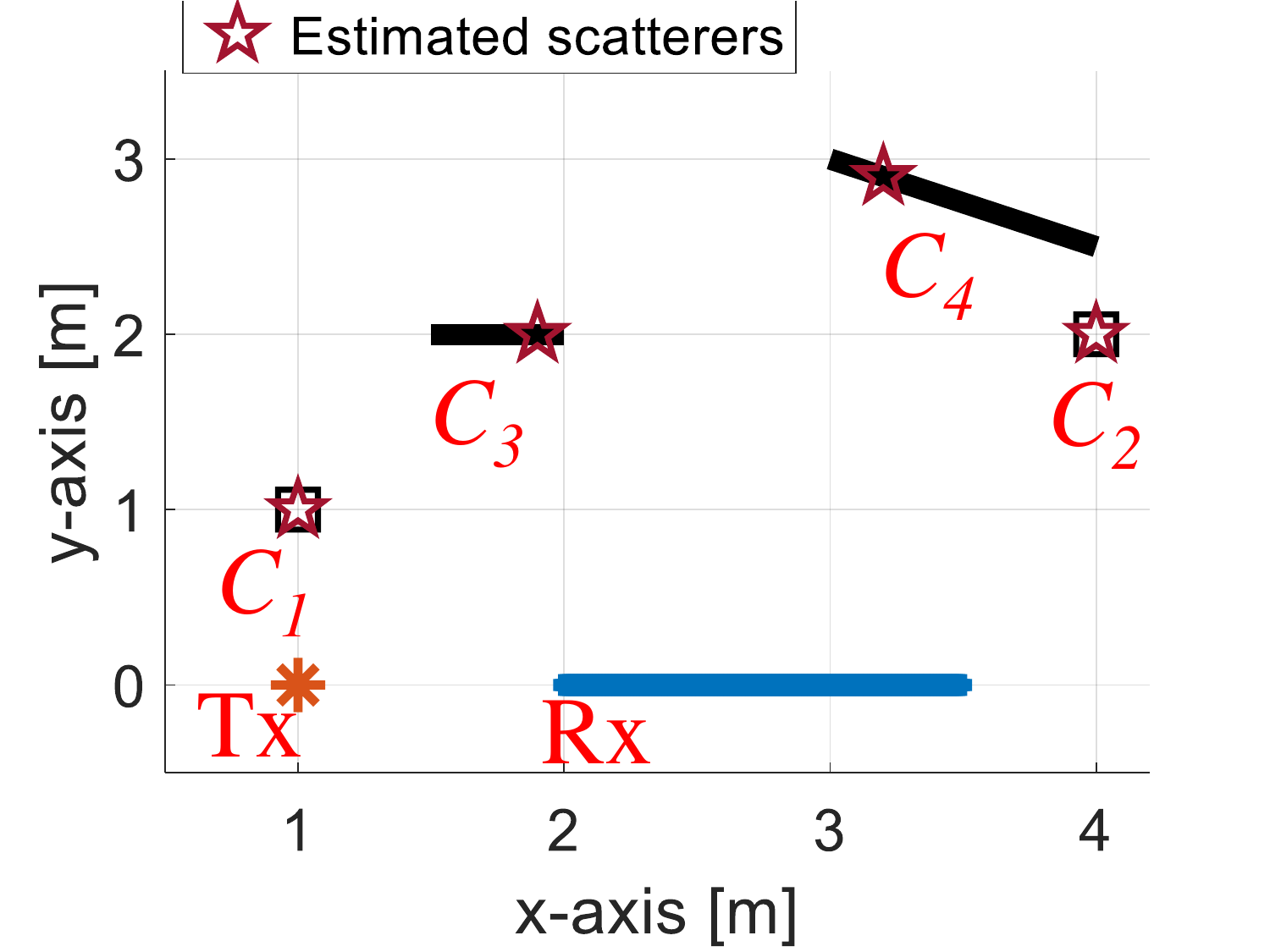}}
    \hfill
    \subfigure[Proposed: GC-SAGE]{\includegraphics[width=0.42\textwidth]{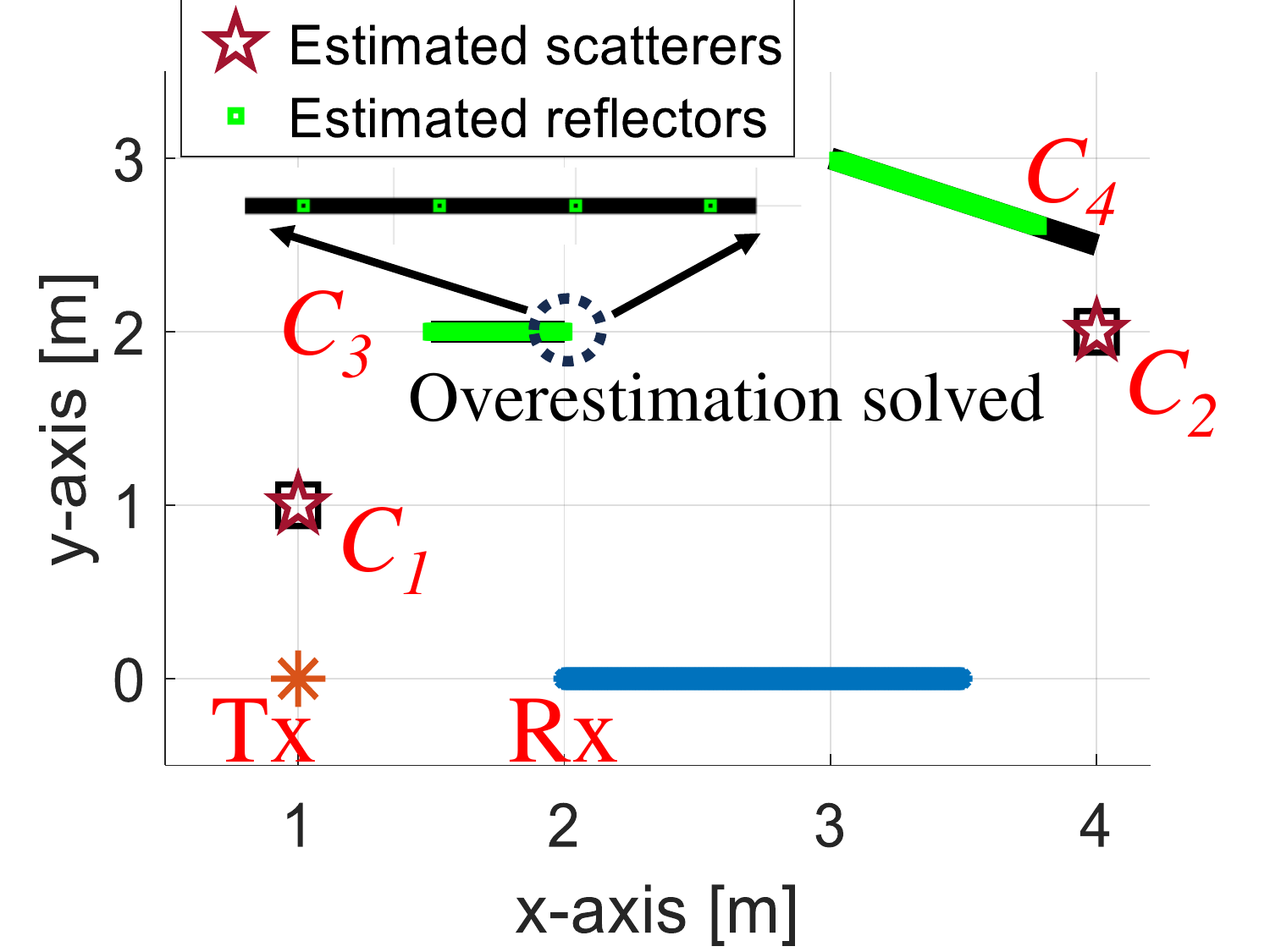}}
    \hfill
    \subfigure[GC-SAGE: SIMO v.s. MIMO]{\includegraphics[width=0.42\textwidth]{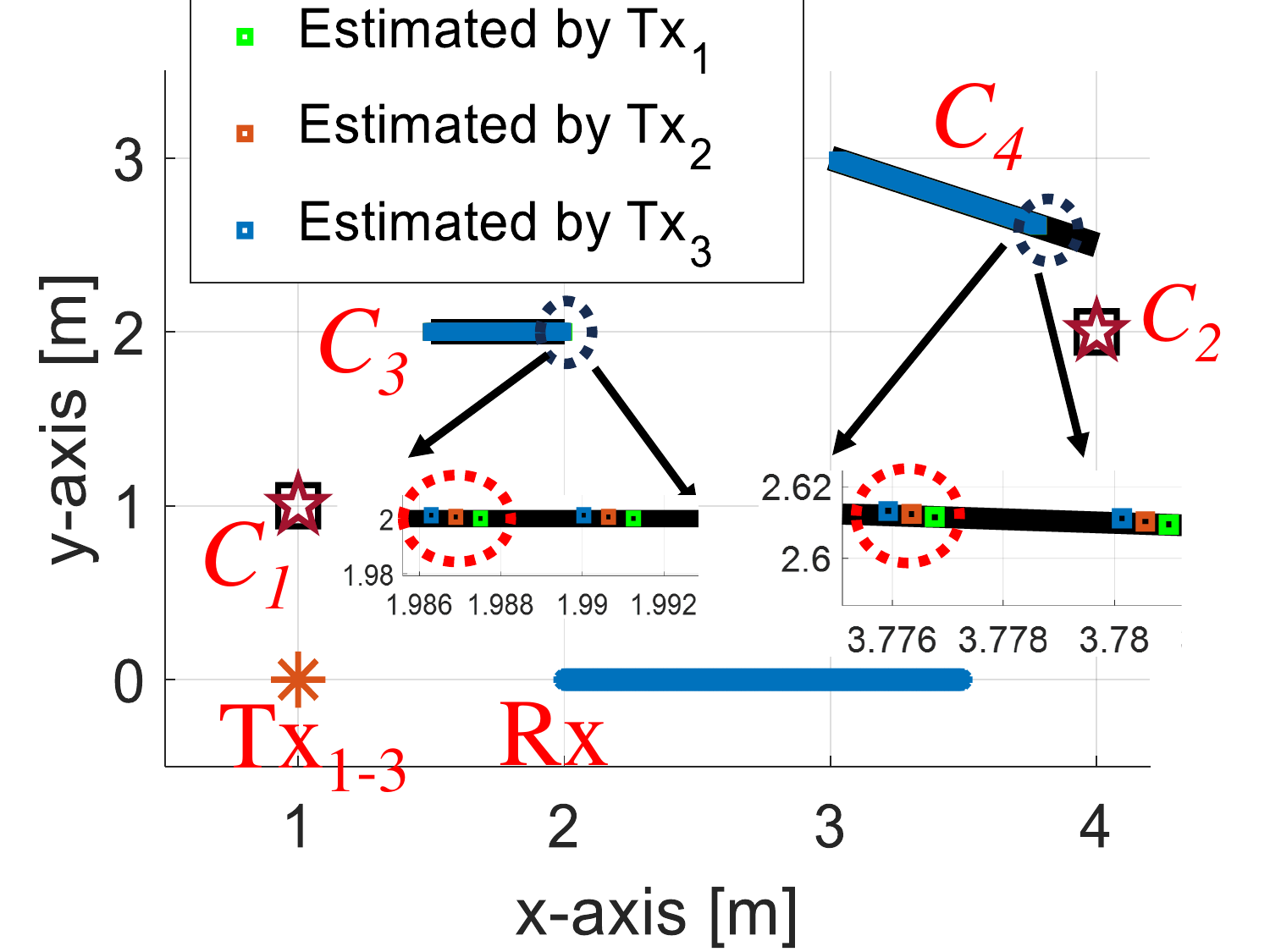}}
    \hfill
    \caption{Simulation case 2: (a) scenario layout and benchmark mapping result; (b) scenario layout and proposed GC-SAGE mapping result; (c) scenario layout and MIMO mapping result.}
    \label{fig:simu_group_1}
    \vspace{-0.5cm}
\end{figure}

\subsection{Simulation Case~2: Hybrid Reflection and Scattering}
\subsubsection{Scenario}
We modify the scenario as shown in Fig.~\ref{fig:simu_group_1} to evaluate the capability of GC-SAGE in hybrid reflection-scattering channels,
where two rough scatterers, denoted by $C_1$ and $C_2$, and two smooth reflecting surfaces, denoted by $C_3$ and $C_4$, are included.
Moreover, to investigate the difference between coherent scattering sources and antenna-dependent specular reflection points, we consider both single-input-multiple-output (SIMO) and MIMO arrays.
The key simulation parameters are summarized in Table~\ref{tab:simu_configs_1}.

%
%

\subsubsection{Estimation and Environment Reconstruction}
%
%
Mapping results of the scattering-only model and the proposed GC-SAGE are shown in Fig.~\ref{fig:simu_group_1}(a) and (b), respectively.
Both approaches can correctly localize the two rough scatterers $C_1$ and $C_2$, while the results of the reflection paths are different.
As shown in Fig.~\ref{fig:simu_group_1}(a), using the scattering-only model, the two smooth reflecting surfaces $C_3$ and $C_4$ are estimated as scatterers.
In contrast, Fig.~\ref{fig:simu_group_1}(b) shows that GC-SAGE correctly identifies both scatterers and reflecting surfaces and provides geometry-consistent localization results.
In particular, the boundary of $C_3$ is well recovered by estimating the SNS coefficient $\gamma_{m,n,l}$ for each reflecting path.
For $C_4$, only partial recovery is obtained because the Rx aperture does not capture all reflected waves from the whole surface.

\subsubsection{SIMO vs. MIMO in Locating Reflecting Surfaces}
The location of the reflecting surface requires each Tx to be sequentially selected as a reference and paired with the Rx array, as explained in \eqref{eq:lines_image}.
Hence, the interaction points may vary for each Tx.
The green, orange, and blue square dots in Fig.~\ref{fig:simu_group_1}(c) correspond to three Tx antennas located at $(1,0)$, $(1.005,0)$, and $(1.01,0)$, respectively.
Although each Tx-Rx pair can localize a set of reflectors, the zoomed-in view shows that the estimated points (green, orange, and blue) are spread across different locations along the same surface.
The MIMO localization results are not coherently combined, which highlights a fundamental difference between MIMO-based localization of point-like scatterers and that of extended specular reflecting surfaces.

\vspace{-0.5cm}
\begin{figure}[t]
\centering
\subfigure[Photo of the field measurement scenario.\label{fig:meas_photo}]{
\begin{tikzpicture}
  \node[anchor=south west,inner sep=0] (img) at (0,0)
    {\includegraphics[width=0.8\linewidth]{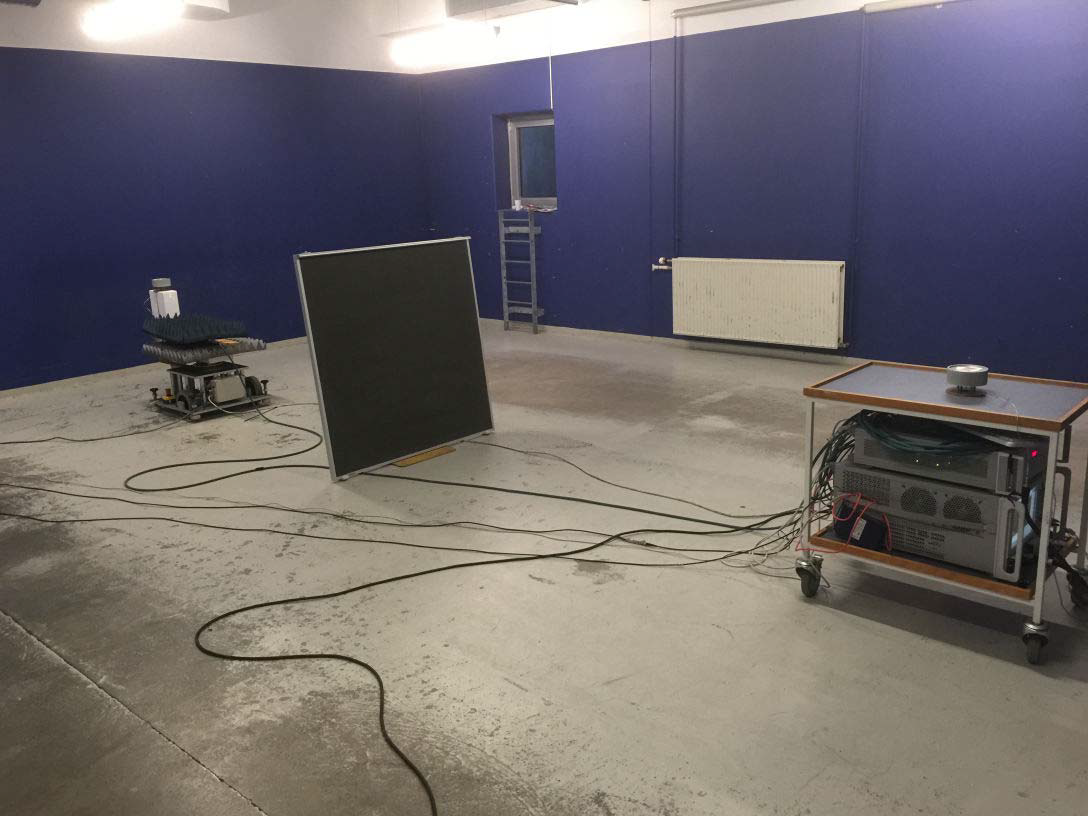}};
  \begin{scope}[x={(img.south east)}, y={(img.north west)}]
    \draw[white,dashed,thick] (0.08,0.42) rectangle (0.26,0.70);
    \draw[white,dashed,thick] (0.28,0.40) rectangle (0.45,0.73);
    \draw[white,dashed,thick] (0.455,0.60) rectangle (0.520,0.88);
    \draw[white,dashed,thick] (0.61,0.57) rectangle (0.780,0.70);
    \draw[white,dashed,thick] (0.70,0.20) rectangle (0.99,0.565);
    \node[white,fill=black,fill opacity=0.35,text opacity=1,inner sep=2pt]
      at (0.16,0.75) {Virtual UCA};
    \node[white,fill=black,fill opacity=0.35,text opacity=1,inner sep=2pt]
      at (0.37,0.55) {Blackboard};
    \node[white,fill=black,fill opacity=0.35,text opacity=1,inner sep=2pt]
      at (0.52,0.90) {Window and metallic stairs};
    \node[white,fill=black,fill opacity=0.35,text opacity=1,inner sep=2pt]
      at (0.69,0.70) {Metallic heater};
    \node[white,fill=black,fill opacity=0.35,text opacity=1,inner sep=2pt]
      at (0.90,0.62) {Tx};
  \end{scope}
\end{tikzpicture}
}
\hfill
 \vspace{-0.5cm}
\subfigure[3D schematic of the measurement geometry.\label{fig:meas_sketch3d}]{
\includegraphics[width=1 \linewidth]{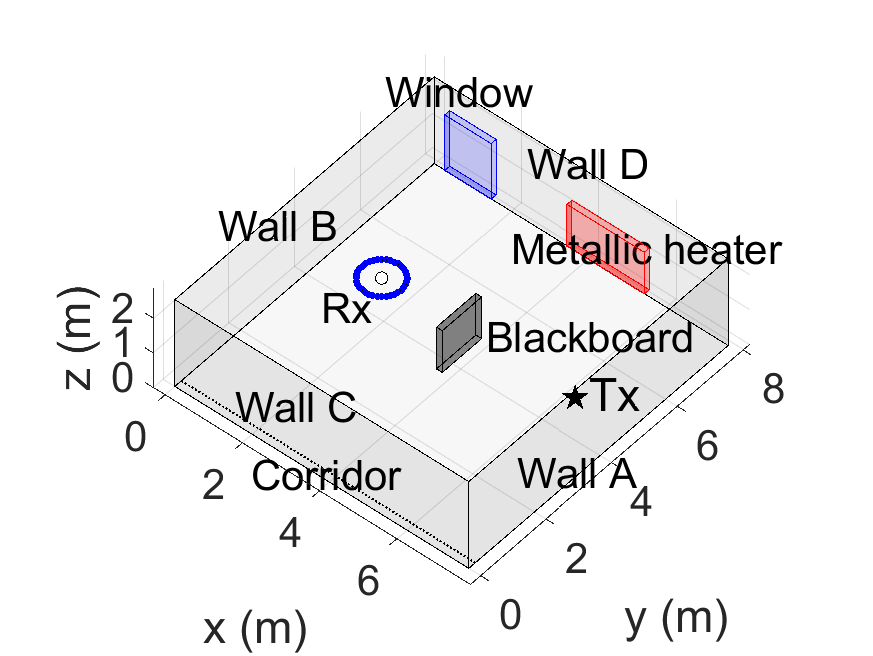}
}
\caption{Measurement environment and geometry used for validation.}
\label{fig:meas_setup}
\end{figure}
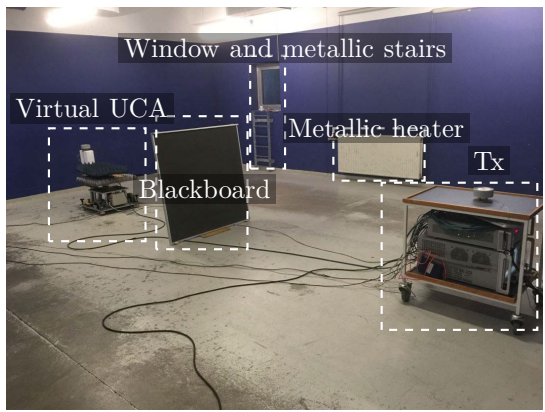
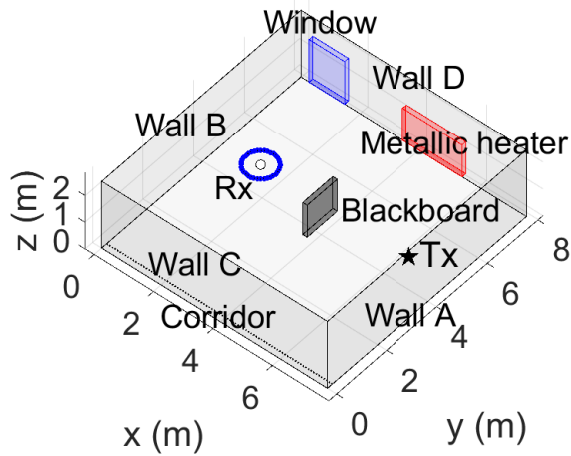
\begin{table}[t]
\centering
\caption{Important Configurations in Field Measurement }
\label{tab:meas_specs}
\begin{tabular}{ll}
\hline
Configurations & Value \\
\hline
Room size~[m$^3$]  & $ 7.7 \times 7.9 \times 2.5 $ \\
Carrier band~[GHz] & $28$-$30$ \\
Bandwidth~[GHz] & $2$ \\
Frequency points & $750$ \\
Tx antenna & Omnidirectional biconical \\
Rx antenna & Omnidirectional biconical \\
Tx height~[m] & $0.84$ \\
Rx height~[m] & $0.84$ \\
UCA radius~[m] & $0.5$ \\
UCA elements (virtual) & $720$ (clockwise rotation) \\
Tx to UCA center distance & $5$~m \\
Scenarios & LoS and OLoS \\
OLoS blocker & $1.2 \,\mathrm{m} \times 1.2 \,\mathrm{m}$ blackboard \\
\hline
\end{tabular}
\end{table}
\begin{figure}[t]
     \centering
          \subfigure[CPDP of the LoS scenario \label{fig:CPDP_LoS}]{
          \begin{tikzpicture}
  \node[inner sep=0] (img) {\includegraphics[width=0.45\textwidth]{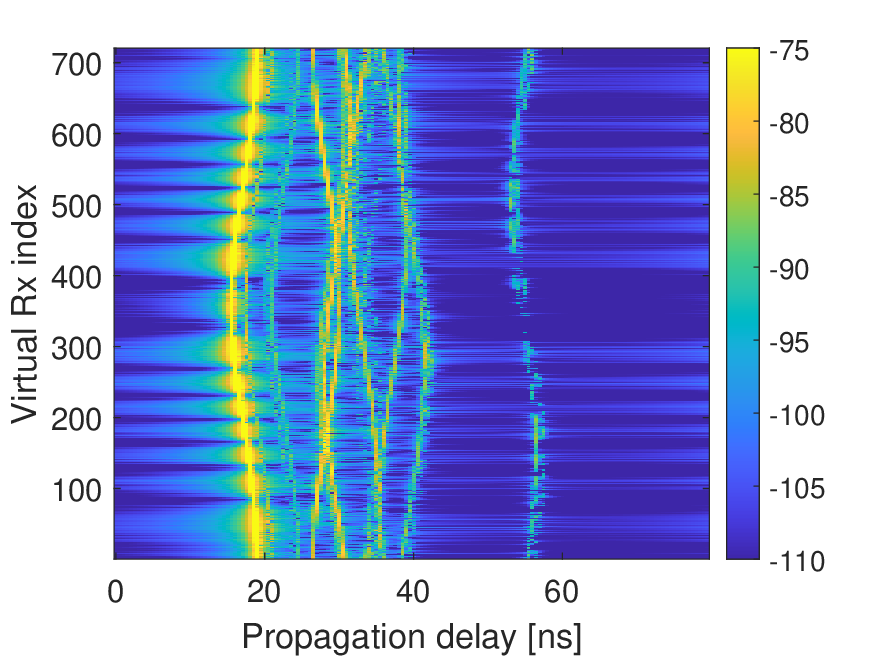}};
  \begin{scope}[shift={(img.south west)}, x={(img.south east)}, y={(img.north west)}]
    \draw[red, line width=1.5pt, ->] (0.285,0.8) -- (0.36,0.8);
    \node[red, font=\bfseries\footnotesize, anchor=south west] at (0.29,0.82) {P 1};
    \draw[red, line width=1.5pt, ->] (0.18,0.83) -- (0.28,0.83);
    \node[red, font=\bfseries\footnotesize, anchor=south west] at (0.18,0.85) {LoS};
    \draw[red, line width=1.5pt, ->] (0.41,0.68) -- (0.51,0.68) ;
    \node[red, font=\bfseries\footnotesize, anchor=south west] at (0.43, 0.69) {P 2};
    \draw[red, line width=1.5pt] (0.42,0.48) circle [radius=0.065];
    \node[red, font=\bfseries\footnotesize, anchor=south west] at (0.38,0.46) {P 3};
  \end{scope}
\end{tikzpicture}
          }
      \hfill
       \vspace{-0.5cm}
    \subfigure[CPDP of the OLoS scenario \label{fig:CPDP_OLoS}]{
          \begin{tikzpicture}
  \node[inner sep=0] (img) {\includegraphics[width=0.45\textwidth]{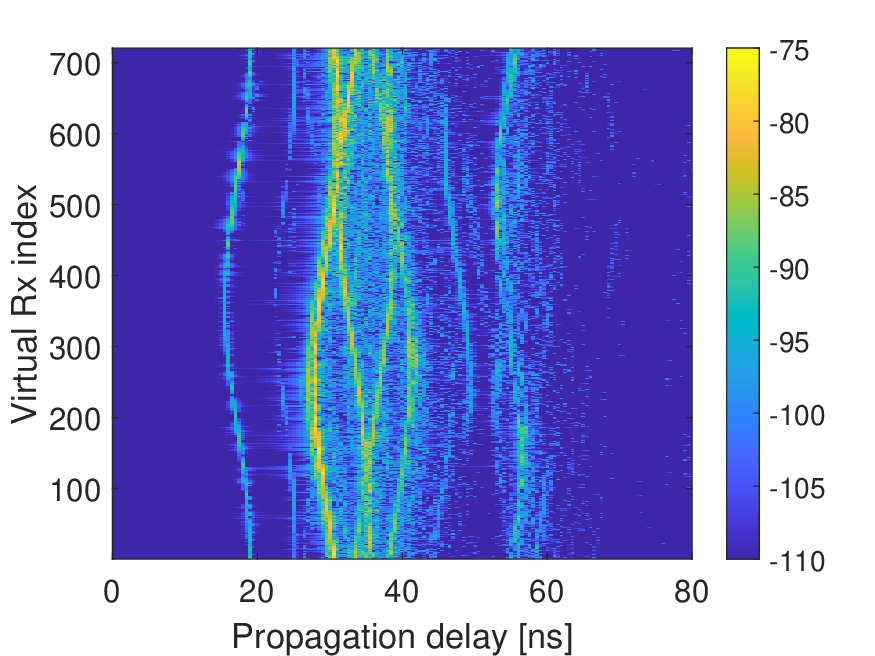}};
  \begin{scope}[shift={(img.south west)}, x={(img.south east)}, y={(img.north west)}]
    \draw[red, line width=1.5pt, ->] (0.285,0.8) -- (0.36,0.8);
    \node[red, font=\bfseries\footnotesize, anchor=south west] at (0.29,0.82) {P 1};
    \draw[red, line width=1.5pt, ->] (0.18,0.83) -- (0.28,0.83);
    \node[red, font=\bfseries\footnotesize, anchor=south west] at (0.18,0.85) {LoS};
    \draw[red, line width=1.5pt, ->] (0.41,0.68) -- (0.51,0.68) ;
    \node[red, font=\bfseries\footnotesize, anchor=south west] at (0.43, 0.69) {P 2};
    \draw[red, line width=1.5pt] (0.41,0.48) circle [radius=0.065];
    \node[red, font=\bfseries\footnotesize, anchor=south west] at (0.38,0.46) {P 3};
  \end{scope}
\end{tikzpicture}
    }
        \caption{CPDPs of both LoS and OLoS measurements.}
    \label{fig:CPDP}
     \vspace{-0.5cm}
\end{figure}
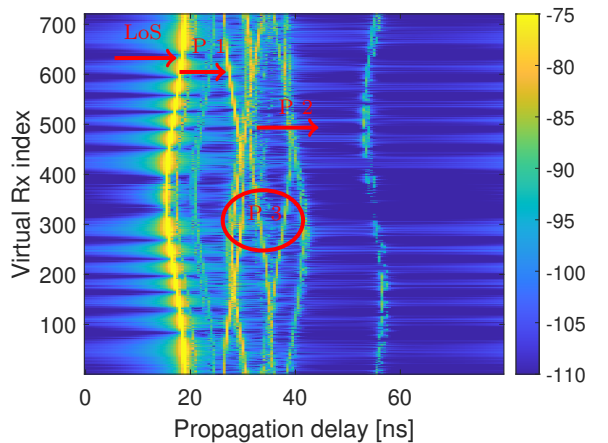
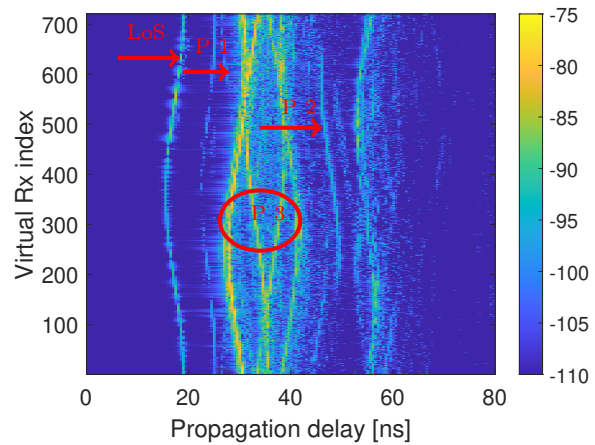
\begin{figure*}[h!]
     \centering
          \subfigure[Estimated scatterers in LoS scenario \label{fig:Scatter_LoS_3D}]{\includegraphics[width=0.48\textwidth]{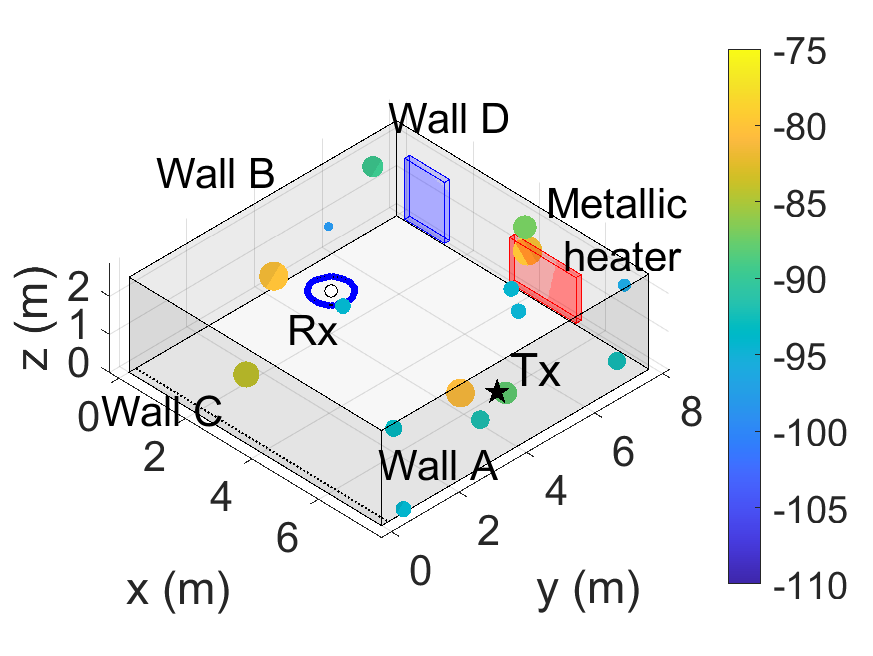}}
      \hfill
    \subfigure[Top view of the estimated scatterers in LoS scenario \label{fig:Scatter_LoS_top}]{\includegraphics[width=0.48\textwidth]{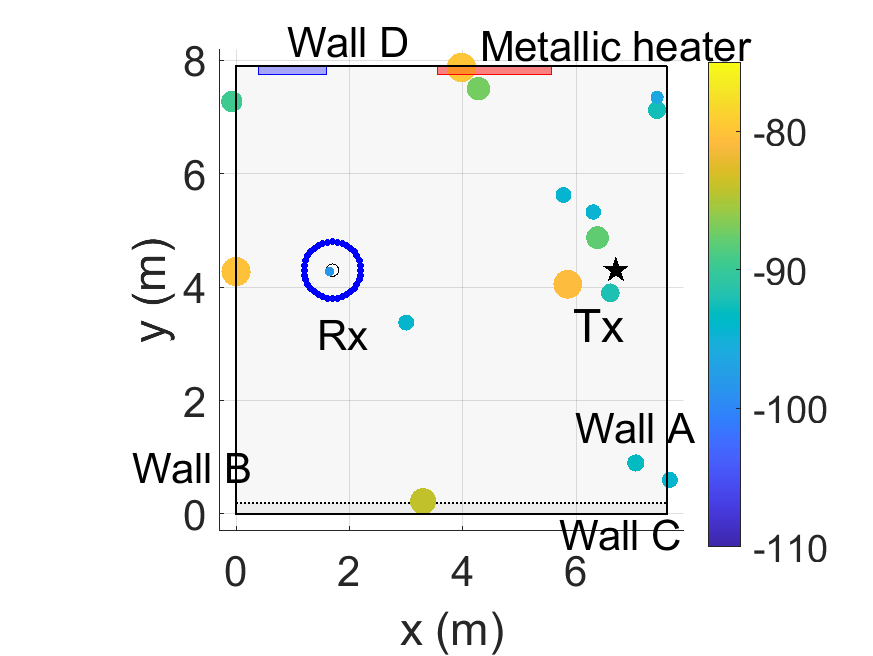}}
    \hfill
     \vspace{-0.2cm}
    \subfigure[Estimated scatterers in OLoS scenario \label{fig:Scatter_OLoS_3D}]{\includegraphics[width=0.48\textwidth]{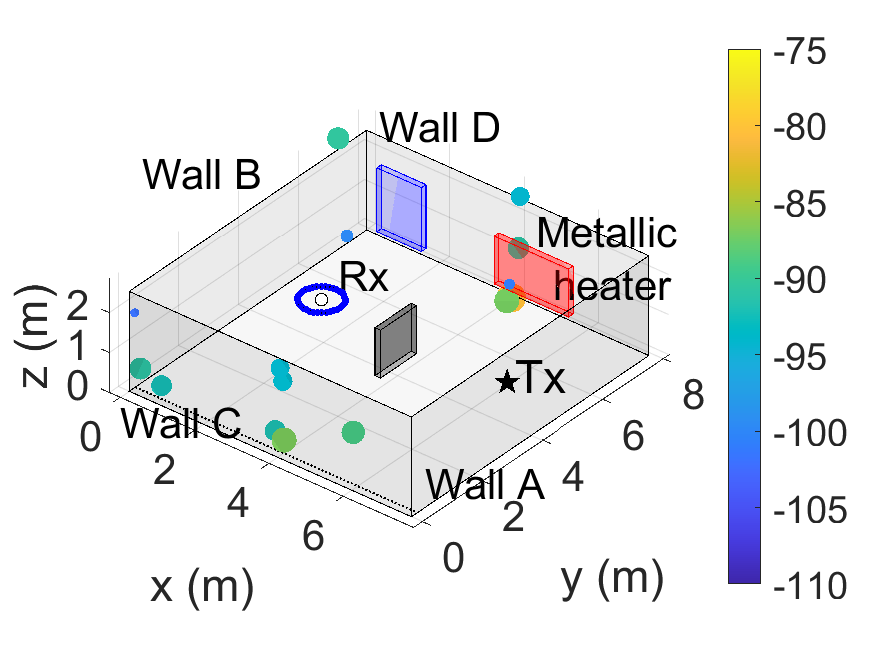}}
      \hfill
    \subfigure[Top view of the estimated scatterers in OLoS scenario \label{fig:Scatter_OLoS_top}]{\includegraphics[width=0.48\textwidth]{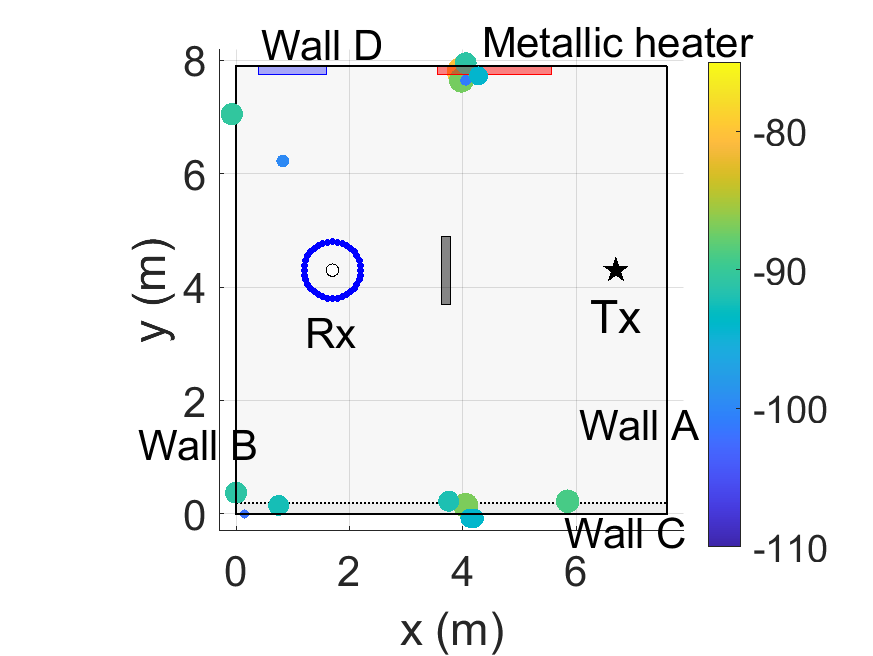}}
    \hfill
     \vspace{-0.2cm}
    \subfigure[Estimated scatterers and reflectors in OLoS scenario \label{fig:Reflect_OLoS_3D}]{
    \begin{tikzpicture}[spy using outlines={rectangle, red, magnification=2, size=1.8cm, connect spies}]
        \node[inner sep=0] (img) {\includegraphics[width=0.48\textwidth]{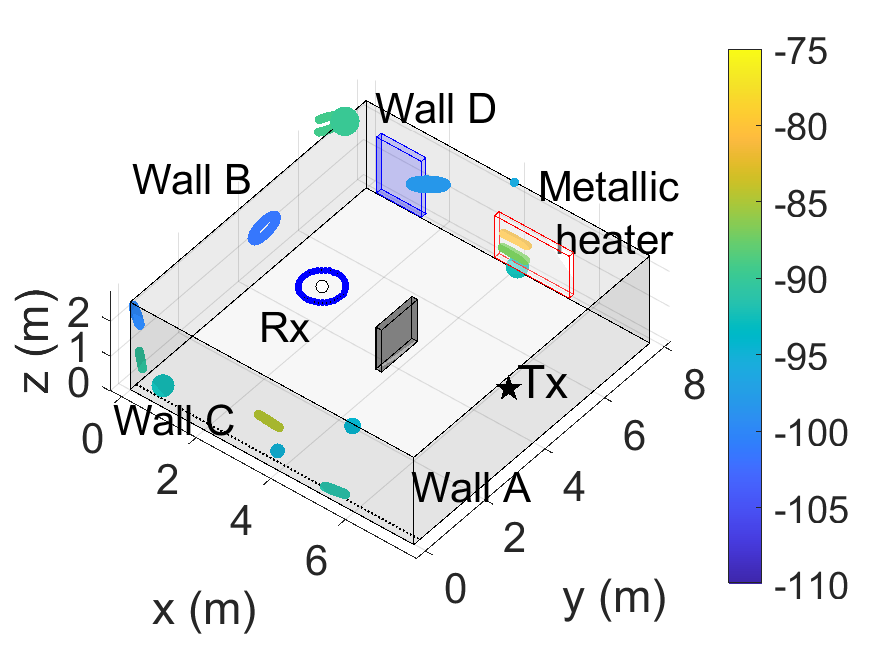}};
        \begin{scope}[shift={(img.south west)}, x={(img.south east)}, y={(img.north west)}]
            \coordinate (heaterCenter) at (0.60,0.62);
            \spy [red] on (heaterCenter) in node [anchor=north east] at (1.5,0);
        \end{scope}
    \end{tikzpicture} }
    \hfill
\subfigure[Top view of the estimated scatterers in OLoS scenario \label{fig:Reflect_OLoS_top}]{\includegraphics[width=0.48\textwidth]{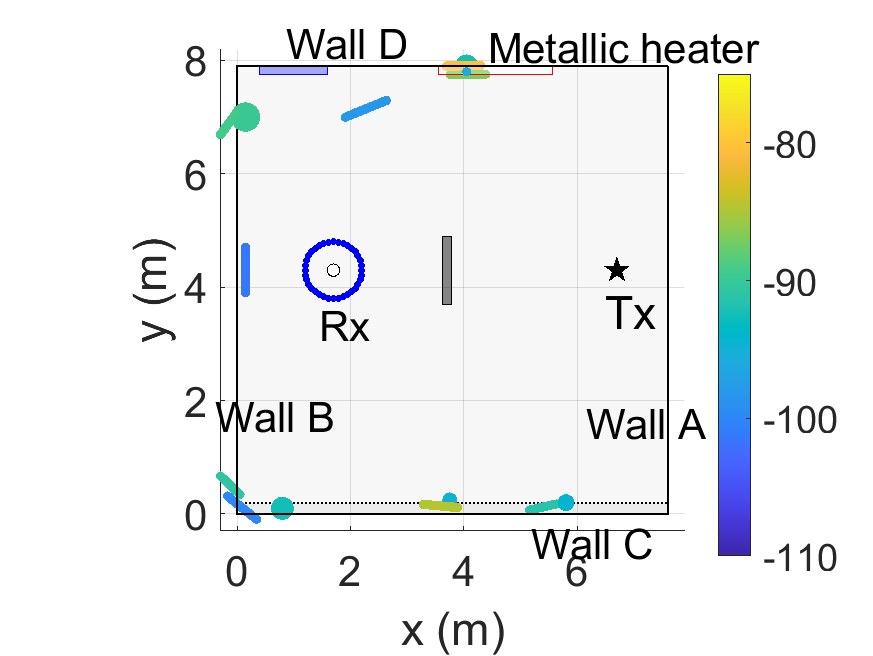}}
    \hfill
        \vspace{-0.2cm}
        \caption{Reconstructed scatterers map in the $3$-D environment: comparison between LoS and OLoS scenarios using proposed and scattering-only consumptions.}
    \label{fig:Compare_blockage}
        \vspace{-0.5cm}
\end{figure*}
\subsection{Field Near-Field Measurement}
%
%
\subsubsection{Measurement Scenario}
%
The measurement campaign was conducted in a basement environment as shown in Fig.~\ref{fig:meas_photo}, where a floor size of approximately $7.7 \,\mathrm{m} \times 7.9 \,\mathrm{m}$ and a few large objects, including metallic stairs and a metallic heater. A vector network analyzer (VNA) recorded the frequency-domain channel response sweeping from $28$ to $30$~GHz with $750$ frequency points. 
An omnidirectional biconical antenna served as the Tx and was mounted on a trolley\footnote{We have to notice that the trolley is not well covered by absorbers, therefore observed scattering/diffraction paths in the LoS scenario measurement.} at a height of $0.84$~m. 
An identical biconical antenna served as the Rx and was placed on a turntable to form a virtual uniform circular array (UCA) by rotating clockwise with radius $0.5$~m and $720$ angular steps\footnote{By forming the virtual UCA, the so-called Rayleigh distance is around $200$~m. Thus, it is reasonable to consider the near-field effects.}. 
The Rx height was kept identical to the Tx height. The distance between the Tx and the UCA center was $5$~m. Two scenarios were considered: a LoS case and an obstructed-LoS (OLoS) case, where a $1.2 \,\mathrm{m} \times 1.2 \,\mathrm{m}$ metallic-substrate blackboard was positioned between Tx and the UCA to block the LoS direction. Additional reference measurements were performed by replacing the Rx biconical antenna with a horn antenna positioned at the center of the UCA. 
The important configurations are listed in Table~\ref{tab:meas_specs}.
More details about the mmWave measurement dataset can be found in \cite{8713575}. 

Accordingly, a $3$-D schematic of the measurement scenarios with main objects and geometry is illustrated in Fig.~\ref{fig:meas_sketch3d}.  
In this $3$-D illustration, the four main walls in the basement are labeled as Wall~A to Wall~D, corresponding to the Tx side, Rx side, corridor side, and metallic-heater side, respectively.
%
Although the picture shows an obstructed blackboard between the Tx and Rx UCA, the measurement was conducted for both LoS and OLoS cases for comparison.

\subsubsection{CPDP Observations}
The concatenated power delay profiles (CPDPs) of the near-field measurements are shown in Fig.~\ref{fig:CPDP}, where Fig.~\ref{fig:CPDP_LoS} and Fig.~\ref{fig:CPDP_OLoS} correspond to the LoS and OLoS cases, respectively. 
To compare the two CPDPs, we added the same set of markers at identical coordinates to highlight the differences between the LoS and OLoS cases.

\textbf{Observations of Blockage}:
The \textbf{LoS} component is clearly visible in both measurements with the shortest delays, while the power is severely reduced in Fig.~\ref{fig:CPDP_OLoS}. 
%
At the same time, some multipath components observed in the LoS scenario disappear in the OLoS case due to partial blockage caused by obstructions. For instance, the path labeled as \textbf{P~1} is clearly visible in Fig.~\ref{fig:CPDP_LoS} but nearly invisible in Fig.~\ref{fig:CPDP_OLoS}, indicating that it is likely associated with reflections from Wall~A or Wall~B, which are obstructed by the obstacle. 

\textbf{Inference on Diffraction Paths}:
The OLoS measurement in Fig.~\ref{fig:CPDP_OLoS} contains additional weak while long-delay paths, which are not observed in the LoS case, e.g., \textbf{P~2}. These paths are likely introduced by edge diffraction from the obstacle (i.e., scattering from the boundary of the obstruction), which then interacts with the surrounding walls, becoming detectable at the receiver. 

\textbf{Consistent Paths in Both LoS and OLoS}:
Some paths remain consistent in both cases. The circled component \textbf{P~3} exhibits a relatively strong and continuous track, with similar delay and spatial trends in both LoS and OLoS results.
This implies that \textbf{P~3} is likely generated by objects, which are less affected by the blockage, such as Wall~C and Wall~D.

\subsubsection{Environment Mapping Comparison with/without Blockage}
%
In the first comparison, we apply the scattering-only model to both the LoS and OLoS measurement data. 
The environment mapping results are shown in Fig.~\ref{fig:Compare_blockage}, where the $3$-D-view and top-view results for the LoS case are in Fig.~\ref{fig:Scatter_LoS_3D} and Fig.~\ref{fig:Scatter_LoS_top}, respectively, while those for the OLoS case are shown in Fig.~\ref{fig:Scatter_OLoS_3D} and Fig.~\ref{fig:Scatter_OLoS_top}. 
At first glance, most estimated scatterers align with the scene geometry and are located on the walls. 
However, there are noticeable differences between the LoS and OLoS cases. 
The path powers of the located scatterers are represented using both colorbar and marker size. 
The subsequent analysis will focus on these strong paths, such as the yellow and green dots, which are more prominent in the channel

\textbf{Effects of Blockage:}
In the LoS case, a strong scatterer is located on Wall~B, while no comparable scatterers are observed on Wall~B in the OLoS result. 
This difference can be attributed to the blocked path $\mathrm{Tx}\!\rightarrow\!\mathrm{Wall~B}\!\rightarrow\!\mathrm{Rx}$ in the OLoS scenario, which is roughly aligned with the \textbf{P~1}\footnote{The delay of \textbf{P~1} in the UCA Rx first increases and then decreases, we can also infer that the signal is reflected from the direction of Wall~B.} in Fig.~\ref{fig:CPDP}. 
Another noticeable difference is that more scatterers are estimated around the Tx in the LoS case, whereas such components are largely absent in the OLoS case. 
As illustrated by the measurement setup in Fig.~\ref{fig:meas_photo}, this region contains several links of the form $\mathrm{Tx}\!\rightarrow\!\mathrm{scatterers}\!\rightarrow\!\mathrm{Rx}$, which are likely blocked by the obstacle in the OLoS case.

\textbf{Strong Paths from Unblocked Walls:}
In both LoS and OLoS cases in Fig.~\ref{fig:Scatter_LoS_top} and Fig.~\ref{fig:Scatter_OLoS_top}, several strong scatterers are consistently located on Wall~C and Wall~D. 
These walls are not blocked by the obstacle, consistent with the measurement observation of \textbf{P~3}, which further validates the algorithm.
Moreover, these dominant scatterers show two features.
First, they cluster around the $x = 4$~m axis, suggesting that strong paths occur when the $\mathrm{Tx} \rightarrow \mathrm{scatterer} \rightarrow \mathrm{Rx}$ geometry approximately satisfies a specular-reflection condition with equal incident and reflection angles. 
Second, high-power scatterers are estimated around the metallic heater, indicating that highly reflective objects contribute to dominant multipath returns. 
These observations motivate us to explicitly distinguish the \textit{reflection} and \textit{scattering} mechanisms.

\subsubsection{Environment Mapping with Hybrid Reflection-Scattering}
We further evaluate the environment mapping performance by considering a \textbf{hybrid reflection-scattering} model in the GC-SAGE. 
The $3$-D and top-view mapping results are shown in Fig.~\ref{fig:Reflect_OLoS_3D} and Fig.~\ref{fig:Reflect_OLoS_top}, respectively. 
Compared with the scattering-only model in Fig.~\ref{fig:Scatter_OLoS_3D} and Fig.~\ref{fig:Scatter_OLoS_top}, most interaction points are reconstructed at similar locations. 
Under the hybrid reflection-scattering model, some previously estimated scatterers are refined as reflection-induced interaction points and are thus reconstructed as reflectors rather than isolated scatterers. 
The difference is most evident around the metallic heater, where several strong components previously mapped as scatterers are now localized as a set of reflectors.%
\footnote{As discussed in \eqref{eq:mstep_2}, the localized set of reflectors is shaped by the Rx array geometry. In this measurement setup, the Rx uses a UCA, and the reconstructed reflector therefore tends to appear as a line-like structure in the azimuth domain.
} 
From the top-view results in Fig.~\ref{fig:Reflect_OLoS_top}, several reconstructed reflector segments coincide with the surrounding walls. This agreement supports the effectiveness of GC-SAGE in localizing both reflectors and scatterers.

\subsubsection{Convergence and Complexity}
%
\begin{figure}[t]
    \centering
\includegraphics[width=0.45\textwidth]{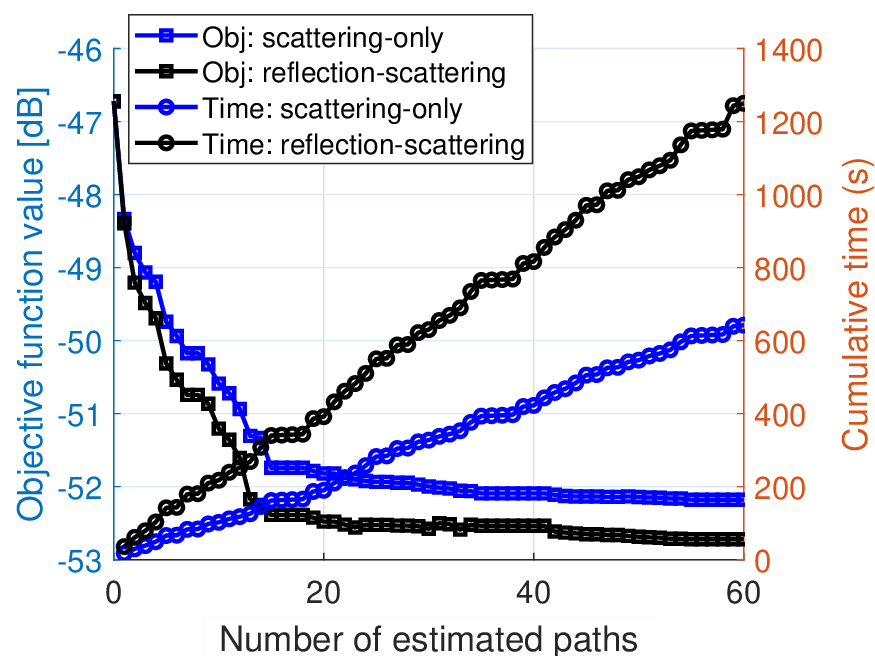}
    \caption{The convergence curves of the objective function and the corresponding time consumption in the OLoS case, using the proposed hybrid reflection-scattering constraints and the scattering-only constraints.  
    }\label{fig:Time_Convergence_OLoS}
     \vspace{-0.5cm}
\end{figure}
Fig.~\ref{fig:Time_Convergence_OLoS} compares the convergence behavior and cumulative runtime of the proposed GC-SAGE under the scattering-only and hybrid reflection-scattering models for the OLoS measurement data. 
From the objective evolution, the hybrid reflection-scattering model converges to a consistently lower objective value than the scattering-only model, indicating a more complete representation of the channel components in realistic scenarios. This improved modeling enables GC-SAGE to better separate and localize reflectors and scatterers, thereby enhancing both channel estimation and environment mapping performance.

%
However, the hybrid model introduces additional computational overhead. As analyzed in Section~\ref{sec:complexity}, the added reflector search and constraint evaluation increase the per-iteration cost, roughly doubling the complexity from $\mathcal{O}\!\left(MN\!\times\!(\eta LV+\eta^2LV^2)\right)$ to $\mathcal{O}\!\left(2MN(\eta LV+\eta^2LV^2)\right)$.
This is also reflected in the cumulative runtime in Fig.~\ref{fig:Time_Convergence_OLoS}, where the hybrid reflection-scattering model roughly doubles the total runtime compared with the scattering-only case.

Overall, both methods converge stably as more paths are iterated. The objective decreases monotonically and then saturates, indicating that the proposed GC-SAGE framework is numerically robust on practical measurement data, even under model mismatch. In comparison, the better-matched hybrid model further reduces the objective and enables more accurate classification and localization of reflectors and scatterers.

\section{Conclusion}\label{sec:conclude}
This paper proposed a GC-SAGE algorithm for near-field channel estimation and environment mapping. In addition to the widely considered scattering model, the SNS phenomena, including blockage, diffraction, and reflection, were modeled within a unified parametric framework based on wave-propagation GCs. The locations of scatterers and reflectors, along with the array geometry, were embedded in this model through the spatial consistency of delay variations across the array. This enabled the GC-SAGE algorithm to iteratively perform joint localization of each path scatterers/reflectors and SNS channel estimation.
Both simulation and measurement-based validations demonstrated that the proposed GC-SAGE algorithm effectively estimated complex SNS near-field channels, along with high-accuracy scatterer/reflector localization and mapping. Notably, for SNS detection, the algorithm overcame the traditional limitations of estimating sparse coefficients by directly estimating the exact amplitude of each path in the channel, thus simplifying the SNS detection process.

There are some interesting observations during the scenario-based validations, which are worth further exploration:
(i)~Scattering sources provide coherent gain in MIMO channels, while reflecting surfaces contribute less coherent gain.
(ii)~Partial blockage may cause certain paths in the channel to transition from near-field to far-field due to the reduction in the effective aperture.
(iii)~SNS is influenced not only by propagation mechanisms but also by wideband effects, e.g., frequency selectivity.

\vspace{-0.5cm}
\bibliographystyle{IEEEtran}
\bibliography{ref.bib}

@ARTICLE{7981398,
  author={Yin, Xuefeng and Wang, Stephen and Zhang, Nan and Ai, Bo},
  journal={IEEE Transactions on Wireless Communications}, 
  title={Scatterer Localization Using Large-Scale Antenna Arrays Based on a Spherical Wave-Front Parametric Model}, 
  year={2017},
  volume={16},
  number={10},
  pages={6543-6556},
  doi={10.1109/TWC.2017.2725260}}

@ARTICLE{9938129,
  author={Hong, Jingxiang and Rodríguez-Pineiro, Jose and Yin, Xuefeng and Yu, Ziming},
  journal={IEEE Transactions on Wireless Communications}, 
  title={Joint Channel Parameter Estimation and Scatterers Localization}, 
  year={2023},
  volume={22},
  number={5},
  pages={3324-3340},
  doi={10.1109/TWC.2022.3217560}}

@ARTICLE{10179246,
  author={Zhou, Zihao and Wang, Cheng-Xiang and Zhang, Li and Huang, Jie and Xin, Lijian and Aggoune, El-HadiM. and Miao, Yang},
  journal={IEEE Transactions on Antennas and Propagation}, 
  title={A Novel {SAGE} Algorithm for Estimating Parameters of Wideband Spatial Nonstationary Wireless Channels With Antenna Polarization}, 
  year={2023},
  volume={71},
  number={9},
  pages={7457-7472},
  doi={10.1109/TAP.2023.3292508}}

@ARTICLE{7501567,
  author={Chen, Jiajing and Wang, Stephen and Yin, Xuefeng},
  journal={IEEE Communications Letters}, 
  title={A Spherical-Wavefront-Based Scatterer Localization Algorithm Using Large-Scale Antenna Arrays}, 
  year={2016},
  volume={20},
  number={9},
  pages={1796-1799},
  doi={10.1109/LCOMM.2016.2585478}}

@ARTICLE{10038714,
  author={Jing, Guangzheng and Hong, Jingxiang and Yin, Xuefeng and Rodriguez-Pineiro, Jose and Yu, Ziming},
  journal={IEEE Transactions on Wireless Communications}, 
  title={Measurement-Based 3-D Channel Modeling With Cluster-of-Scatterers Estimated Under Spherical-Wave Assumption}, 
  year={2023},
  volume={22},
  number={9},
  pages={5828-5843},
  doi={10.1109/TWC.2023.3237564}}

@ARTICLE{9215972,
  author={Wymeersch, Henk and He, Jiguang and Denis, Benoit and Clemente, Antonio and Juntti, Markku},
  journal={IEEE Vehicular Technology Magazine}, 
  title={Radio Localization and Mapping With Reconfigurable Intelligent Surfaces: Challenges, Opportunities, and Research Directions}, 
  year={2020},
  volume={15},
  number={4},
  pages={52-61},
  doi={10.1109/MVT.2020.3023682}}

@ARTICLE{witrisal2016high,
  author={K. Witrisal and P. Meissner and E. Leitinger and Y. Shen and C. Gustafson and F. Tufvesson and K. Haneda and D. Dardari and A. F. Molisch and A. Conti and M. Z. Win},
  title={High-accuracy localization for assisted living: {5G} systems will turn multipath channels from foe to friend},
  journal={IEEE Signal Process. Mag.},
  volume={33},
  number={2},
  pages={59-70},
  year={2016},
  doi={10.1109/MSP.2015.2504328}
}

@article{ling2017experimental,
  title={Experimental characterization and multipath cluster modeling for 13-17 {GHz} indoor propagation channels},
  author={Ling, Cen and Yin, Xuefeng and Wang, Haowen and Thom{\"a}, Reiner S},
  journal={IEEE Transactions on antennas and propagation},
  volume={65},
  number={12},
  pages={6549-6561},
  year={2017},
  publisher={IEEE}
}

@article{feder1988parameter,
  title={Parameter estimation of superimposed signals using the {EM} algorithm},
  author={Feder, Meir and Weinstein, Ehud},
  journal={IEEE Transactions on acoustics, speech, and signal processing},
  volume={36},
  number={4},
  pages={477-489},
  year={1988},
  publisher={IEEE}
}

@article{fleury1999channel,
  title={Channel parameter estimation in mobile radio environments using the {SAGE} algorithm},
  author={Fleury, Bernard H and Tschudin, Martin and Heddergott, Ralf and Dahlhaus, Dirk and Pedersen, K Ingeman},
  journal={IEEE Journal on selected areas in communications},
  volume={17},
  number={3},
  pages={434-450},
  year={1999},
  publisher={IEEE}
}

@inproceedings{liu2024debris,
  title={Debris Sensing Based on {Leo} Constellation: An Intersatellite Channel Parameter Estimation Approach},
  author={Liu, Yuan and Shankar, MR Bhavani and Wu, Linlong and Ottersten, Bj{\"o}rn},
  booktitle={ICASSP 2024-2024 IEEE International Conference on Acoustics, Speech and Signal Processing (ICASSP)},
  pages={13171-13175},
  year={2024},
  organization={IEEE}
}

@ARTICLE{8713575,
  author={Cai, Xuesong and Fan, Wei},
  journal={IEEE Transactions on Communications}, 
  title={A Complexity-Efficient High Resolution Propagation Parameter Estimation Algorithm for Ultra-Wideband Large-Scale Uniform Circular Array}, 
  year={2019},
  volume={67},
  number={8},
  pages={5862-5874},
  doi={10.1109/TCOMM.2019.2916700}}

@ARTICLE{10559769,
  author={Liu, Yuan and Ahmadi, Moein and Fuchs, Johann and Alaee-Kerahroodi, Mohammad and Bhavani Shankar, M. R.},
  journal={IEEE Transactions on Antennas and Propagation}, 
  title={Dynamic Indoor mmWave MIMO Radar Simulation: An Image Rendering-Based Approach}, 
  year={2025},
  volume={73},
  number={4},
  pages={1984-1999},
  doi={10.1109/TAP.2024.3412269}}

@ARTICLE{Yuan_TWC25,
  author={Liu, Yuan and Wu, Linlong and Cai, Xuesong and Shankar, M. R. Bhavani},
  journal={IEEE Transactions on Wireless Communications}, 
  title={Graph-Based Multi-Bounce Modeling and Channel Parameter Estimation for Indoor Sensing}, 
  year={2025},
  volume={24},
  number={5},
  pages={4219-4234},
  doi={10.1109/TWC.2025.3537081}}

@ARTICLE{10445208,
  author={Han, Chong and Chen, Yuhang and Yan, Longfei and Chen, Zhi and Dai, Linglong},
  journal={IEEE Wireless Communications}, 
  title={Cross Far- and Near-Field Wireless Communications in Terahertz Ultra-Large Antenna Array Systems}, 
  year={2024},
  volume={31},
  number={3},
  pages={148-154},
  doi={10.1109/MWC.003.2300004}}

@INPROCEEDINGS{9827865,
  author={Liu, Yuan and Wu, Linlong and Alaee-Kerahroodi, Mohammad and R, Bhavani Shankar Mysore},
  booktitle={2022 IEEE 12th Sensor Array and Multichannel Signal Processing Workshop (SAM)}, 
  title={A {3D} Indoor Localization Approach Based on Spherical Wave-front and Channel Spatial Geometry}, 
  year={2022},
  volume={},
  number={},
  pages={101-105},
  doi={10.1109/SAM53842.2022.9827865}}

@ARTICLE{8642926,
  author={Dokhanchi, Sayed Hossein and Mysore, Bhavani Shankar and Mishra, Kumar Vijay and Ottersten, Björn},
  journal={IEEE Transactions on Aerospace and Electronic Systems}, 
  title={A mmWave Automotive Joint Radar-Communications System}, 
  year={2019},
  volume={55},
  number={3},
  pages={1241-1260},
  doi={10.1109/TAES.2019.2899797}}

@book{bertoni2001radio,
  author    = {H. L. Bertoni},
  title     = {Radio Propagation for Modern Wireless Systems},
  year      = {2001},
  publisher = {Prentice Hall},
  pages     = {107-140},
  note      = {Chapters 4-5 focus on diffraction and scattering}
}

@article{new_look_1974,
  author    = {M. H. C. Weng and M. P. O'Rourke},
  title     = {A new look at the statistical model identification},
  journal   = {IEEE Trans. Autom. Control},
  volume    = {AC-19},
  number    = {6},
  pages     = {716-723},
  year      = {1974},
  month     = {Dec.}
}

@book{mishra2024signal,
  author    = {Mishra, Kumar Vijay and Shankar, MR Bhavani and Ottersten, Bjorn and Swindlehurst, A Lee},
  title     = {Signal processing for joint radar communications},
  publisher = {John Wiley \& Sons},
  year      = {April 2024},
  edition   = { },
  volume    = { },
  series    = { },
  address   = { },
  isbn      = {978-1-119-79555-1}
}

@ARTICLE{8828030,
  author={Mishra, Kumar Vijay and Bhavani Shankar, M.R. and Koivunen, Visa and Ottersten, Bjorn and Vorobyov, Sergiy A.},
  journal={IEEE Signal Processing Magazine}, 
  title={Toward Millimeter-Wave Joint Radar Communications: A Signal Processing Perspective}, 
  year={2019},
  volume={36},
  number={5},
  pages={100-114},
  doi={10.1109/MSP.2019.2913173}}

@INPROCEEDINGS{liu2025environment,
   author={Liu, Yuan and Wu, Linlong and Cai, Xuesong and Shankar, M. R. Bhavani},
  booktitle={2025 33rd European Signal Processing Conference (EUSIPCO)}, 
  title={Environment Reconstruction in Multi-Bounce Channels with Array Partial Blockage}, 
  year={2025},
  volume={},
  number={},
  pages={1208-1212},
  doi={10.23919/EUSIPCO63237.2025.11226715}}

@INPROCEEDINGS{10447918,
  author={Huang, Huiping and Zhang, Tianjian and Yin, Feng and Liao, Bin and Wymeersch, Henk},
  booktitle={ICASSP 2024 - 2024 IEEE International Conference on Acoustics, Speech and Signal Processing (ICASSP)}, 
  title={Joint {DOA} Estimation and Distorted Sensor Detection Under Entangled Low-Rank and Row-Sparse Constraints}, 
  year={2024},
  volume={},
  number={},
  pages={12851-12855},
  doi={10.1109/ICASSP48485.2024.10447918}}

@ARTICLE{9940939,
  author={Yuan, Zhiqiang and Zhang, Jianhua and Ji, Yilin and Pedersen, Gert Frølund and Fan, Wei},
  journal={IEEE Transactions on Antennas and Propagation}, 
  title={Spatial Non-Stationary Near-Field Channel Modeling and Validation for Massive {MIMO} Systems}, 
  year={2023},
  volume={71},
  number={1},
  pages={921-933},
  doi={10.1109/TAP.2022.3218759}}

@ARTICLE{1510955,
  author={Jeng-Shiann Jiang and Ingram, M.A.},
  journal={IEEE Transactions on Communications}, 
  title={Spherical-wave model for short-range {MIMO}}, 
  year={2005},
  volume={53},
  number={9},
  pages={1534-1541},
  doi={10.1109/TCOMM.2005.852842}}

@article{lu2024tutorial,
  author  = {Haiquan Lu and Yong Zeng and Changsheng You and Yu Han and Jiayi Zhang and Zhe Wang and Zhenjun Dong and Shi Jin and Cheng-Xiang Wang and Tao Jiang and Xiaohu You and Rui Zhang},
  title   = {A Tutorial on Near-Field {XL-MIMO} Communications Toward {6G}},
  journal = {IEEE Communications Surveys \& Tutorials},
  volume  = {26},
  number  = {4},
  pages   = {2213--2257},
  year    = {2024},
  doi     = {10.1109/COMST.2024.3387749}
}

@article{channelestimationxlmimo,
  author  = {Mingyao Cui and Linglong Dai},
  title   = {Channel Estimation for Extremely Large-Scale {MIMO}: Far-Field or Near-Field?},
  journal = {IEEE Transactions on Communications},
  volume  = {70},
  number  = {4},
  pages   = {2663--2677},
  year    = {2022}
}

@INPROCEEDINGS{1351040,
  author={Thoma, R.S. and Landmann, M. and Sommerkorn, G. and Richter, A.},
  booktitle={Proceedings of the 21st IEEE Instrumentation and Measurement Technology Conference (IEEE Cat. No.04CH37510)}, 
  title={Multidimensional high-resolution channel sounding in mobile radio}, 
  year={2004},
  volume={1},
  number={},
  pages={257-262 Vol.1},
  doi={10.1109/IMTC.2004.1351040}}

@ARTICLE{leitinger2023datafusion,
  author={Leitinger, Erik and Venus, Alexander and Teague, Bryan and Meyer, Florian},
  journal={IEEE Transactions on Signal Processing}, 
  title={Data Fusion for Multipath-Based SLAM: Combining Information From Multiple Propagation Paths}, 
  year={2023},
  volume={71},
  number={},
  pages={4011-4028},
  doi={10.1109/TSP.2023.3310360}}

@INPROCEEDINGS{10694028,
  author={Kaltiokallio, Ossi and Ge, Yu and Talvitie, Jukka and Rastorgueva-Foi, Elizaveta and Wymeersch, Henk and Valkama, Mikko},
  booktitle={2024 IEEE 25th International Workshop on Signal Processing Advances in Wireless Communications (SPAWC)}, 
  title={Bistatic mmWave Mapping in Obstructed Environments Using Double-bounce Signals}, 
  year={2024},
  volume={},
  number={},
  pages={106-110},
  doi={10.1109/SPAWC60668.2024.10694028}}

@article{feng2024multipathghost_trs,
  author  = {Feng, Ruoyu and De Greef, Eddy and Rykunov, Maxim and Pollin, Sofie and Bourdoux, Andr{\'e} and Sahli, Hichem},
  title   = {Multipath Ghost Recognition and Joint Target Tracking With Wall Estimation for Indoor {MIMO} Radar},
  journal = {IEEE Transactions on Radar Systems},
  volume  = {2},
  pages   = {154--164},
  year    = {2024},
  doi     = {10.1109/TRS.2024.3354509}
}

@article{duggal2025diffraction,
  author  = {Duggal, Gaurav and Buehrer, R. Michael and Dhillon, Harpreet S. and Reed, Jeffrey H.},
  title   = {Diffraction-Aided Wireless Positioning},
  journal = {IEEE Transactions on Wireless Communications},
  year    = {2025},
  doi     = {10.1109/TWC.2025.3537959}
}

@article{JHZhang2026jsac_fr3_xlmimo,
  author  = {Xu, Huixin and Zhang, Jianhua and Tang, Pan and Xing, Hongbo and Miao, Haiyang and Zhang, Nan and Li, Jian and Wu, Jianming and Yang, Wenfei and Zhang, Zhening and Jiang, Wei and He, Zijian and Haghighat, Afshin and Wang, Qixing and Liu, Guangyi},
  journal = {IEEE Journal on Selected Areas in Communications},
  title   = {Near-Field Propagation and Spatial Non-Stationarity Channel Model for 6--24 {GHz} ({FR3}) Extremely Large-Scale {MIMO}: Adopted by 3{GPP} for {6G}},
  year    = {2026},
  pages   = {1-1},
  doi     = {10.1109/JSAC.2026.3650888}
}

@ARTICLE{9755276,
  author={Wei, Zhongxiang and Liu, Fan and Masouros, Christos and Su, Nanchi and Petropulu, Athina P.},
  journal={IEEE Communications Magazine}, 
  title={Toward Multi-Functional {6G} Wireless Networks: Integrating Sensing, Communication, and Security}, 
  year={2022},
  volume={60},
  number={4},
  pages={65-71},
  doi={10.1109/MCOM.002.2100972}}

@article{liu2025doppler,
  title        = {Doppler Robust Vortex Wavefront Design for Integrated Sensing and Communication},
  author       = {Liu, Yuan and Long, Wen-Xuan and Bhavani Shankar, M. R. and Moretti, Marco and Chen, Rui and Ottersten, Bj{\"o}rn},
  journal      = {arXiv preprint arXiv:2512.03802},
  year         = {2025},
  eprint       = {2512.03802},
  archivePrefix= {arXiv},
  primaryClass = {eess.SP},
  doi          = {10.48550/arXiv.2512.03802}
}

@ARTICLE{11113304,
  author={Long, Wen-Xuan and Song, Wenfei and Liu, Yuan and Liu, Yongjun and Moretti, Marco and Chen, Rui},
  journal={IEEE Open Journal of Vehicular Technology}, 
  title={{GPS}-Denied {ISAC} Vehicle Localization Based on {mmWave} Radar and Identification}, 
  year={2025},
  volume={6},
  number={},
  pages={2343-2357},
  doi={10.1109/OJVT.2025.3596143}}

@ARTICLE{11126933,
  author  = {Rodr{\'\i}guez-Pi{\~n}eiro, Jos{\'e} and Wei, Zhongxiang and Wang, Jingjing and Guti{\'e}rrez, Carlos A. and Correia, Luis M.},
  journal = {IEEE Open Journal of Vehicular Technology},
  title   = {{6G}-Enabled Vehicle-to-Everything Communications: Current Research Trends and Open Challenges},
  year    = {2025},
  volume  = {6},
  pages   = {2358-2391},
  doi     = {10.1109/OJVT.2025.3599570}
}

@ARTICLE{11183598,
  author={Long, Wen-Xuan and Moretti, Marco and Bacci, Giacomo and Sanguinetti, Luca},
  journal={IEEE Wireless Communications Letters}, 
  title={Near-Field {MMSE} Channel Estimation for {THz} {RIS}-Aided Communications With Electromagnetic Interference}, 
  year={2025},
  volume={14},
  number={12},
  pages={4152-4156},
  doi={10.1109/LWC.2025.3615392}}

@ARTICLE{Liu2012COST2100,
  author  = {Liu, Lingfeng and Oestges, Claude and Poutanen, Juho and Haneda, Katsuyuki and Vainikainen, Pertti and Quitin, Fran{\c{c}}ois and Tufvesson, Fredrik and De Doncker, Philippe},
  journal = {IEEE Wireless Communications},
  title   = {The COST 2100 MIMO Channel Model},
  year    = {2012},
  volume  = {19},
  number  = {6},
  pages   = {92-99},
  doi     = {10.1109/MWC.2012.6393523}
}

@ARTICLE{Wu2015NonStationaryMassiveMIMO,
  author  = {Wu, Shangbin and Wang, Cheng-Xiang and Haas, Harald and Aggoune, el-Hadi M. and Alwakeel, Mohammed M. and Ai, Bo},
  journal = {IEEE Transactions on Wireless Communications},
  title   = {A Non-Stationary Wideband Channel Model for Massive MIMO Communication Systems},
  year    = {2015},
  volume  = {14},
  number  = {3},
  pages   = {1434-1446},
  doi     = {10.1109/TWC.2014.2366153}
}

@ARTICLE{Tropp2007OMP,
  author  = {Tropp, Joel A. and Gilbert, Anna C.},
  journal = {IEEE Transactions on Information Theory},
  title   = {Signal Recovery From Random Measurements Via Orthogonal Matching Pursuit},
  year    = {2007},
  volume  = {53},
  number  = {12},
  pages   = {4655-4666},
  doi     = {10.1109/TIT.2007.909108}
}

@ARTICLE{10964143,
  author={Zhang, Peng and Chen, Yuan and Du, Jianhe and Li, Xingwang and Yang, Gang and Yuen, Chau},
  journal={IEEE Transactions on Vehicular Technology}, 
  title={Channel Parameter Estimation and Localization for Near-Field {XL}-{MIMO} Communications}, 
  year={2025},
  volume={74},
  number={9},
  pages={14781-14786},
  doi={10.1109/TVT.2025.3560492}}

@ARTICLE{9598863,
  author={Wei, Xiuhong and Dai, Linglong},
  journal={IEEE Communications Letters}, 
  title={Channel Estimation for Extremely Large-Scale Massive {MIMO}: Far-Field, Near-Field, or Hybrid-Field?}, 
  year={2022},
  volume={26},
  number={1},
  pages={177-181},
  doi={10.1109/LCOMM.2021.3124927}}

@article{Wei2025FundamentalLimitsNFPartII,
  title        = {Fundamental Limits for Near-Field Sensing - Part II: Wide-Band Systems},
  author       = {Wei, Tong and Mishra, Kumar Vijay and Bhavani Shankar, M. R. and Ottersten, Bj{\"o}rn},
  journal      = {arXiv preprint arXiv:2512.24962},
  year         = {2025},
  eprint       = {2512.24962},
  archivePrefix= {arXiv},
  primaryClass = {eess.SP},
  doi          = {10.48550/arXiv.2512.24962}
}

@INPROCEEDINGS{10942642,
  author    = {Xu, Jiaqi and Ottersten, Bj\"{o}rn and Swindlehurst, A. Lee},
  booktitle = {2024 58th Asilomar Conference on Signals, Systems, and Computers},
  title     = {Partially-Blocked Near-Field Sensing: Joint Source DoA and Blockage Range Estimation},
  year      = {2024},
  pages     = {1871--1875},
  doi       = {10.1109/IEEECONF60004.2024.10942642}
}

@ARTICLE{10509715,
  author={Tang, Anzheng and Wang, Jun-Bo and Pan, Yijin and Zhang, Wence and Zhang, Xiaodan and Chen, Yijian and Yu, Hongkang and de Lamare, Rodrigo C.},
  journal={IEEE Transactions on Communications}, 
  title={Joint Visibility Region and Channel Estimation for Extremely Large-Scale {MIMO} Systems}, 
  year={2024},
  volume={72},
  number={10},
  pages={6087-6101},
  doi={10.1109/TCOMM.2024.3394757}}
\end{document}